\documentclass[journal]{IEEEtran}

\usepackage{tabstackengine}
\usepackage{cite}
\usepackage{graphicx}
\usepackage{floatrow}
\usepackage{amssymb} 
\usepackage{multirow}
\usepackage{caption}
\usepackage{xcolor}
\usepackage{authblk}
\usepackage{indentfirst}
\usepackage{physics}
\usepackage{mathtools}
\usepackage{amsmath}
\usepackage{subfig}
\usepackage{fancyhdr}
\usepackage{gensymb}
\usepackage{amsmath,mleftright}
\usepackage{xparse}
\usepackage{booktabs}
\usepackage{float}

\NewDocumentCommand{\evalat}{sO{\big}mm}{%
  \IfBooleanTF{#1}
   {\mleft. #3 \mright|_{#4}}
   {#3#2|_{#4}}%
}
\usepackage{amsmath}
\usepackage{caption}
\usepackage{fixltx2e}
\begin{document}
%
\title{Large-Scale Crosstalk-Corrected Thermo-Optic Phase Shifter Arrays in Silicon Photonics}
%
%
%

\author{
B. Volkan Gurses,~\IEEEmembership{Student Member,~IEEE,} Reza Fatemi,~\IEEEmembership{Member,~IEEE,} Aroutin Khachaturian,~\IEEEmembership{Member,~IEEE,} and~Ali Hajimiri,~\IEEEmembership{Fellow,~IEEE}
\thanks{Manuscript received February 01, 2022; revised June 04, 2022; accepted June 13, 2022. (\textit{Corresponding author: B. Volkan Gurses}.)}
\thanks{
B. Volkan Gurses, Reza Fatemi, Aroutin Khachaturian, and Ali Hajimiri are with the Department of Electrical Engineering, California Institute of Technology, Pasadena, CA, 91125 USA (e-mail: gurses@caltech.edu; sfatemi@caltech.edu; akhachat@caltech.edu; hajimiri@caltech.edu).
}}

%
%

\markboth{IEEE Journal of Selected Topics in Quantum Electronics}%
{Shell \MakeLowercase{\textit{Gurses et al.}}: 
Large-Scale Thermo-Optic Phase Shifter Arrays with Independent Phase Control}
%



\maketitle

\begin{abstract}
We introduce a thermo-optic phase shifter (TOPS) array architecture with independent phase control of each phase shifter for large-scale and high-density photonic integrated circuits with two different control schemes: pulse amplitude modulation (PAM) and pulse width modulation (PWM). We realize a compact spiral TOPS and a 288-element high-density row-column TOPS array with this architecture and drive TOPS with waveforms of both control schemes and of different array sizes. We present a thermal excitation model and a finite difference method-based simulation to simulate large-scale TOPS arrays and compare both schemes experimentally and theoretically. We also analyze the effects of thermal crosstalk in the realized TOPS array and implement a thermal crosstalk correction algorithm with the developed model. The high-density TOPS array architecture and the thermal crosstalk correction algorithm pave the way for high-density TOPS arrays with independent phase control in large-scale photonic integrated circuits interfaced with electronics limited in voltage swing and bandwidth.
\end{abstract}

\begin{IEEEkeywords}
optical phase shifters, thermo-optic effects, time division multiplexing, crosstalk, large-scale circuits, integrated optoelectronics, silicon
\end{IEEEkeywords}

%
\IEEEpeerreviewmaketitle


\section{Introduction}
%
%
%
%
\IEEEPARstart{I}{ntegrated} photonics allows numerous bulk optic components such as lenses, modulators, and fiber-optics to be integrated on a thin substrate to realize millimeter-scale, energy-efficient optical systems with better manufacturability and cost \cite{Miller1969}. The advent of large-scale photonic integrated circuits (PICs) largely driven by silicon photonics enabled high-performing systems driving practical applications in sensing, communications, and computing \cite{Thomson2016}.\par
Silicon photonics leverage the high refractive index of silicon to enable the realization of optical waveguides with tight mode confinement. This allows compact nanophotonic structures to be implemented on chip at a high volume. Combined with its CMOS compatibility, silicon photonics enable the largest scale PICs with tens of thousands of components realized on a single chip \cite{Sun2013}. Consequently, many large-scale systems such as optical phased arrays \cite{Fatemi2019,Khachaturian2021,Khachaturian2021b}, programmable photonic circuits \cite{Bogaerts2020}, optical neural networks \cite{Shen2017}, and quantum photonic processors \cite{Wang2018} have been implemented on silicon photonics platform.\par
Large-scale PICs ubiquitously utilize phase shifters for modulation, tuning, calibration of systematic phase errors, and correction of random phases along the signal path \cite{Yang2015}. As the circuits scale, a larger number of phase shifters are needed for precise signal manipulation. Thermo-optic phase shifters (TOPS) are a good candidate for high-density silicon PICs because of their low optical loss and small form factor \cite{Watts2013, Harris2014}. However, high-density TOPS arrays suffer from thermal crosstalk \cite{Jacques2019}.\par
In high-density integration of TOPS, unless photonics and electronics are both monolithically integrated, electrical connections required to drive each phase shifter in a direct addressing scheme can pose a problem for scaling. Furthermore, if each phase shifter requires its own driver, the complexity of the control electronics and their power consumption scale rapidly \cite{Miller2020}, becoming a limiting factor for scalability.\par
\begin{figure}[h!]%
    \centering
    \includegraphics[width=0.85\textwidth]{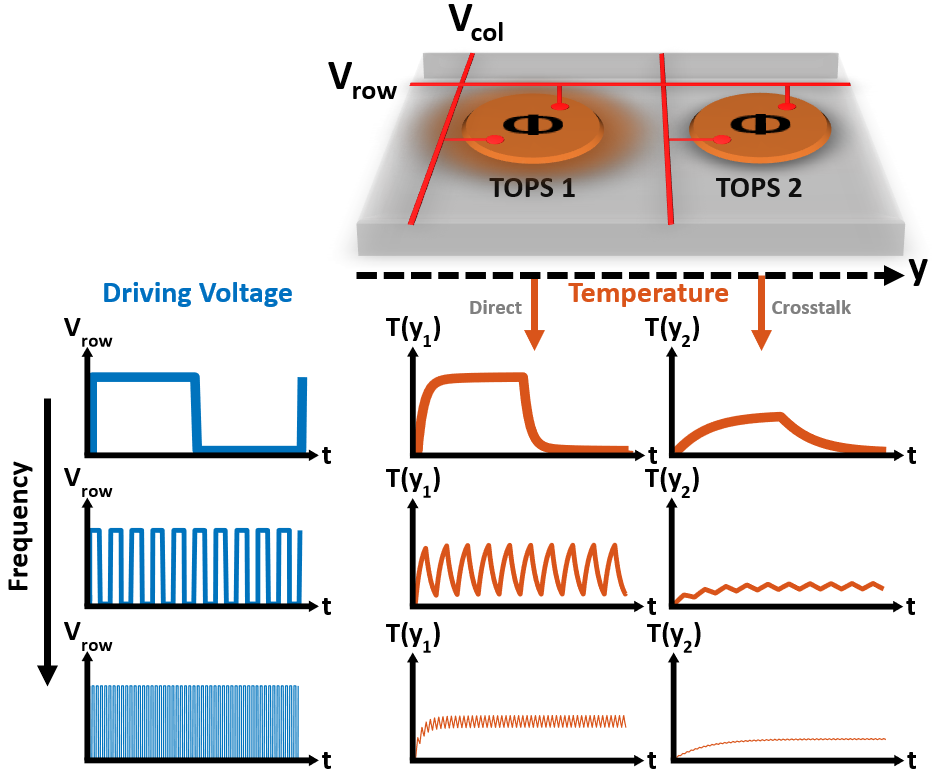}
    \caption{Temporal multiplexing of TOPS arrays showing the AC averaging for direct modulation and thermal crosstalk.}
    \label{fig:concept}
\end{figure}

To overcome these challenges, row-column or matrix addressing schemes have been implemented \cite{Fatemi2019, Khachaturian2021, Khachaturian2021b, Ribeiro2020}. In these schemes, the number of electrical connections scales as $N$ with $N^2$ phase shifters. This reduces the complexity of the electronic interface as opposed to a direct addressing scheme where each phase shifter has a separate connection and where the number of electrical connections scales as $N$. In the row-column architecture, each phase shifter is connected between a column and row node in series with a diode. The diode serves as a switch for the current to flow between a column and row node and modulate the phase shifter. In such a scheme, columns determine which diodes will be in forward bias, while rows determine the amplitude of the electrical signal. Electrical signals used to modulate the TOPS are time multiplexed across multiple columns, by which each TOPS receives an average electrical power proportional to the duty cycle as shown in Fig. \ref{fig:concept}. The average electrical power received by each TOPS can then be controlled by either pulse-amplitude modulation (PAM) or pulse-width modulation (PWM) of the row voltages.\par
In this paper, we realize a compact TOPS design and a large-scale 288-element row-column TOPS array with independent phase control using a folded row-column architecture \cite{Fatemi2019} along with integrated Mach-Zehnder interferometers (MZIs) and row-column photodiodes (PDs) to probe the TOPS phase shifts and calibrate the array.
Driving a TOPS+MZI test structure with PAM and PWM waveforms of different array sizes, we compare PAM and PWM driving schemes. We introduce a thermal excitation model accompanied with a finite difference method-based simulation to predict the experimental results and analyze both schemes. We also showcase the effects of thermal crosstalk between phase shifters and implement a thermal crosstalk correction algorithm to completely cancel the crosstalk in the row-column TOPS array.
\section{Thermo-Optic Phase Shifter} \label{label1}
TOPS utilizes the waveguide refractive index dependence on temperature. The phase shift, $\Phi$, due to refractive index, $n$, is
\begin{equation}
    \Phi=2\pi L\frac{n}{\lambda_0}
\end{equation}
where $\lambda_0$ is the wavelength in free space and $L$ is the length of the heated waveguide. Then, the phase shift in the TOPS is 
\begin{equation}
    \Delta \Phi=\frac{2\pi L}{\lambda_0}\Delta n\\
    =\frac{2\pi L}{\lambda_0}\gamma\Delta T
\end{equation}
where $\Delta T$ is the temperature difference, and $\gamma$ is the temperature coefficient of refractive index. Then, $\Delta T_{\pi}$, temperature difference for $\pi$ phase shift is
\begin{equation}
    \Delta T_{\pi}=\frac{\lambda_0}{2L\gamma}
    \label{eq:Tpi}
\end{equation}
A single TOPS can be modeled as a simple RC circuit, with $P_{\pi}$, power required for $\pi$ phase shift, defined as
\begin{equation}
    P_{\pi}=Ak\Delta T_{\pi}=Ak\frac{\lambda_0}{2L\gamma}
    \label{eq:Ppi}
\end{equation}
where $A$ is the effective surface area and $k$ is the effective thermal conductivity between the waveguide and the heat sink. In the RC model, $G=1/R=Ak$ is the thermal conductance of TOPS to the heat sink.\par
Another parameter to characterize a TOPS is its time constant, $\tau$, similarly defined with the first-order RC approximation as
\begin{equation}
    \tau=\frac{C}{Ak}
    \label{eq:tau}
\end{equation}
where $C$ is the heat capacity of TOPS.
\subsection{Thermal Excitation Model}
\begin{figure}[h!]%
    \centering
    \includegraphics[width=0.95\textwidth]{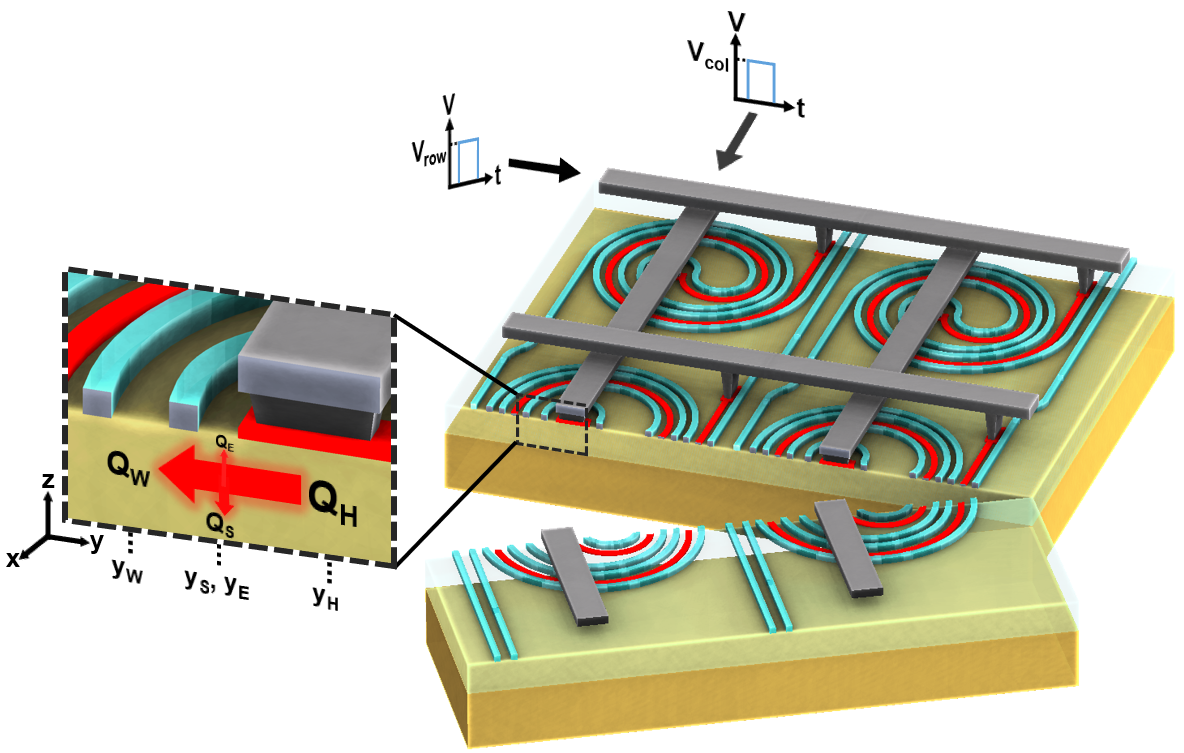}
    \caption{Cross-section of TOPS array showing the interactions between the four main structures.}
    \label{fig:heatmodel}
\end{figure}
We now develop a time-varying lumped model starting with a 3D analysis of TOPS seen in Fig. \ref{fig:heatmodel}, which shows the cross-section of the realized TOPS design and array. Without loss of generality, we consider a heater on one side of the waveguide in an XYZ space. Here, we analyze the dynamics between four structures: heater, waveguide, substrate, and the environment acting as a perfect heat sink. We assume uniform material for all structures, uniform temperature distribution within each structure and in X and Z directions. We define a time-varying temperature function, $T(y,t)$, that characterizes the local instantaneous temperature. When some instantaneous electrical power, $P(t)$, is applied to the resistive heater, this power is expended through four mechanisms (ignoring power dissipated in forms other than Joule heating):
\begin{itemize}
    \item Heat used to increase the heater temperature ($q_H=\rho_Hc_Hd T(y,t)d V$) where $\rho_H$, $c_H$ and $V$ are the heater mass density, specific heat capacity and volume.) 
    \item Heat exchange with the waveguide ($q_{W}=-k_{WH}d td \vec{A}\cdot \nabla T(y,t)$ where $k_{WH}$ is the effective thermal conductivity between the waveguide and the heater, $d\vec{A}$ is the differential surface area vector, and $\nabla T(y,t)$ is the instantaneous temperature gradient.)
    \item Heat exchange with the substrate ($q_{S}=-k_{SH}d td \vec{A}\cdot \nabla T(y,t)$ where $k_{SH}$ is the effective thermal conductivity between the substrate and the heater.)
    \item Heat exchange with surrounding environment other than the waveguide and substrate ($q_{E}=-k_{EH}d td \vec{A}\cdot \nabla T(y,t)$ where $k_{EH}$ is the effective thermal conductivity between the environment and the heater.)
\end{itemize}
\subsubsection{No Crosstalk}
First, we neglect the crosstalk between phase shifters. 
Then, $dP(t)dt=q_H+q_{W}+q_S+q_E$. Going through the derivation in the Supplementary Material, we have the matrix ODE for the lumped model of the system.
\begin{align}
    \frac{d}{dt}
    \begin{bmatrix}
    T_H\\
    T_W\\
    T_S
    \end{bmatrix}
    =
    \begin{bmatrix}
    -\frac{D_H}{C_H} & \frac{G_{WH}}{C_H} & \frac{G_{SH}}{C_H} \\
    \frac{G_{WH}}{C_W} & -\frac{D_W}{C_W} & \frac{G_{SW}}{C_W}\\
    \frac{G_{SH}}{C_S} & \frac{G_{SW}}{C_S} & -\frac{D_S}{C_S}
    \end{bmatrix}
    \begin{bmatrix}
    T_H\\
    T_W\\
    T_S
    \end{bmatrix}
    +
    \begin{bmatrix}
    \frac{P+G_{EH}T_E}{C_H}\\
    \frac{G_{EW}T_E}{C_W}\\
    \frac{G_{ES}T_E}{C_S}
    \end{bmatrix}
    \label{eq:matrixode}
\end{align}
where we defined $G_{ij}=1/R_{ij}=A_{ij}k_{ij}$. Since we assumed uniform temperature distribution within each structure, $k_{ij}$ reduces to effective thermal contact conductivity between the respective structures. Furthermore, $D_H=G_{WH}+G_{SH}+G_{EH}$, $D_W=G_{WH}+G_{SW}+G_{EW}$, and $D_S=G_{SH}+G_{SW}+G_{ES}$ are the sum of thermal conductances for the matrix diagonal elements. Note that $T_E$ is a constant as the environment acts as a heat sink. \par
We also take note that substrate usually facilitates the thermal conduction between structures. Therefore,
\begin{equation}
\begin{aligned}
    R_{EW}&=R_{ES}+R_{SW}\\
    R_{EH}&=R_{ES}+R_{SH}\\
    R_{WH}&=R_{SW}+R_{SH}
\end{aligned}
\end{equation}
\subsubsection{With Crosstalk}
Now, we include the crosstalk between phase shifters and repeat the analysis (see Supplementary Material) by scaling the single TOPS model to N phase shifters. Hence, we have the generalized matrix ODE.\par
\begin{align}
    \frac{d}{dt}
    \begin{bmatrix}
    \textbf{T}_{H}\\
    \textbf{T}_{W}\\
    \textbf{T}_{S}\\
    \end{bmatrix}
    =
    \textbf{M}
    \begin{bmatrix}
    \textbf{T}_{H}\\
    \textbf{T}_{W}\\
    \textbf{T}_{S}\\
    \end{bmatrix}
    +
    \begin{bmatrix}
    \textbf{P}+\textbf{G}_{EH}T_E\\
    \textbf{G}_{EW}T_E\\
    \textbf{G}_{ES}T_E
    \end{bmatrix}
    \label{eq:matrixodefull}
\end{align}
where $\textbf{T}_{X}=\begin{bmatrix}
T_{X,1}\\
\vdots\\
T_{X,N}
\end{bmatrix}$, $\textbf{P}=\begin{bmatrix}
\frac{P_1}{C_{H,1}}\\
\vdots\\
\frac{P_N}{C_{H,N}}
\end{bmatrix}$, $\textbf{G}_{EX}=\begin{bmatrix}
\frac{G_{EX,1}}{C_{X,1}}\\
\vdots\\
\frac{G_{EX,N}}{C_{X,N}}
\end{bmatrix}$, and\footnote{We define $D_{H,i}=S_{H,iH}+S_{WH,i}+S_{SH,i}+G_{EH,i}$, $D_{W,i}=S_{W,iH}+S_{W,iW}+S_{SW,i}+G_{EW,i}$, $D_{S,i}=S_{S,iH}+S_{S,iW}+S_{S,iS}+G_{ES,i}$, where $S_{H,iH}=\sum_{j=1}^N
    \left(G_{H,iH,j}\right)-G_{H,iH,i}$, $S_{WH,i}=\sum_{j=1}^N
    \left(G_{W,jH,i}\right)$, $S_{SH,i}=\sum_{j=1}^N
    \left(G_{S,jH,i}\right)$, $S_{W,iH}=\sum_{j=1}^N
    \left(G_{W,iH,j}\right)$, $S_{W,iW}=\sum_{j=1}^N
    \left(G_{W,iW,j}\right)-G_{W,iW,i}$, $S_{SW,i}=\sum_{j=1}^N
    \left(G_{S,jW,i}\right)$, $S_{S,iH}=\sum_{j=1}^N
    \left(G_{S,iH,j}\right)$, $S_{S,iW}=\sum_{j=1}^N
    \left(G_{S,iW,j}\right)$, $S_{S,iS}=\sum_{j=1}^N
    \left(G_{S,iS,j}\right)-G_{S,i,S,i}$.} $\textbf{M}=$
\begin{align}
\hspace{-5pt}
\begingroup
\setlength\arraycolsep{0pt}
    \begin{bmatrix}
    -\frac{D_{H,1}}{C_{H,1}} & \hdots & \frac{G_{H,1H,N}}{C_{H,1}} & \frac{G_{W,1H,1}}{C_{H,1}} & \hdots & \frac{G_{W,NH,1}}{C_{H,1}} & \frac{G_{S,1H,1}}{C_{H,1}} & \hdots & \frac{G_{S,NH,1}}{C_{H,1}}\\
    \vdots & \ddots & \vdots & \vdots & \ddots & \vdots & \vdots & \ddots & \vdots\\
    \frac{G_{H,1H,N}}{C_{H,N}} & \hdots & -\frac{D_{H,N}}{C_{H,N}} & \frac{G_{W,1H,N}}{C_{H,N}} & \hdots & \frac{G_{W,NH,N}}{C_{H,N}} & \frac{G_{S,1H,N}}{C_{H,N}} & \hdots & \frac{G_{S,NH,N}}{C_{H,N}}\\
    \frac{G_{W,1H,1}}{C_{W,1}} & \hdots & \frac{G_{W,1H,N}}{C_{W,1}} & -\frac{D_{W,1}}{C_{W,1}} & \hdots & \frac{G_{W,1W,N}}{C_{W,1}} & \frac{G_{S,1W,1}}{C_{W,1}} & \hdots & \frac{G_{S,NW,1}}{C_{W,1}}\\
    \vdots & \ddots & \vdots & \vdots & \ddots & \vdots & \vdots & \ddots & \vdots\\
    \frac{G_{W,NH,1}}{C_{W,N}} & \hdots & \frac{G_{W,NH,N}}{C_{W,N}} & \frac{G_{W,1W,N}}{C_{W,N}} & \hdots & -\frac{D_{W,N}}{C_{W,N}} & \frac{G_{S,1W,N}}{C_{W,N}} & \hdots & \frac{G_{S,NW,N}}{C_{W,N}}\\
    \frac{G_{S,1H,1}}{C_{S,1}} & \hdots & \frac{G_{S,1H,N}}{C_{S,1}} & \frac{G_{S,1W,1}}{C_{S,1}} & \hdots & \frac{G_{S,1W,N}}{C_{S,1}} & -\frac{D_{S,1}}{C_{S,1}} & \hdots & \frac{G_{S,1S,N}}{C_{S,1}}\\
    \vdots & \ddots & \vdots & \vdots & \ddots & \vdots & \vdots & \ddots & \vdots\\
    \frac{G_{S,NH,1}}{C_{S,N}} & \hdots & \frac{G_{S,NH,N}}{C_{S,N}} & \frac{G_{S,NW,1}}{C_{S,N}} & \hdots & \frac{G_{S,NW,N}}{C_{S,N}} & \frac{G_{S,NS,1}}{C_{S,N}} & \hdots & -\frac{D_{S,N}}{C_{S,N}}
    \end{bmatrix}
\endgroup
\end{align}
We also again take note that substrate facilitates the thermal conduction between structures. Therefore, for each TOPS,
\begin{equation}
\begin{aligned}
    R_{EW,i}&=R_{ES,i}+R_{S,iW,i}\\
    R_{EH,i}&=R_{ES,i}+R_{S,iH,i}\\
    R_{W,iH,i}&=R_{S,iH,i}+R_{S,iW,i}
\end{aligned}
\end{equation}
This also means crosstalk between TOPS is entirely due to $G_{S,iS,j}=1/R_{S,iS,j}$. Hence,
\begin{equation}
\begin{aligned}
    R_{H,iH,j}&=R_{S,iH,i}+R_{S,jH,j}+R_{S,iS,j}\\
    R_{W,iW,j}&=R_{S,iW,i}+R_{S,jW,j}+R_{S,iS,j}\\
    R_{W,iH,j}&=R_{S,iH,i}+R_{S,jW,j}+R_{S,iS,j}\\
    R_{S,iH,j}&=R_{S,jH,j}+R_{S,iS,j}\\
    R_{S,iW,j}&=R_{S,jW,j}+R_{S,iS,j}\\
\end{aligned}
\end{equation}
For any array of phase shifters, this system of ODEs can be solved numerically or through variation of parameters to obtain the time-domain response of the heaters, waveguides, and substrate. This enables fast modeling of large-scale TOPS arrays without the need for computationally-intensive FDTD simulations enabling efficient engineering of driving waveforms for arbitrary system parameters.
\subsection{Finite Difference Simulation}
Using the framework developed with the thermal excitation model, we implement a finite difference method-based time-domain simulation, numerically solving the aforementioned matrix ODEs to simulate large-scale TOPS arrays. Electrical signals to drive the TOPS array are defined by $P_i$. Therefore, we construct an ideal signal and pass it through an FIR low pass filter designed to simulate the bandwidth of the electronics. Then, to reduce the matrix size and decrease computation time in the simulation, we make some simplifications to the model. We make a first-order approximation for the frequency range we are characterizing the array. We set the substrate parameters, $C_S$ and $G_{ES}$ significantly high, making the substrate an ideal heat sink. Since substrate heating introduces a low-frequency pole to the system, this simplification makes the simulation at low frequencies less accurate but can be removed from the simulation. Furthermore, we define the heater to have low $C_H$, which is a valid assumption for our design as will be shown by the first-order frequency response in our characterization. However, for different designs, this simplification can also be modified. For thermal crosstalk simulations, we assume only first-order crosstalk, meaning $G_{S,iS,i\pm n}=G_{SS}^{(n)}=0$ with $n>1$. We finally approximate $G=G_{SW}=G_{SH}$. With these simplifications, we rederive (\ref{eq:Tpi}), (\ref{eq:Ppi}), and (\ref{eq:tau}) to find the remaining simulation parameters, namely $C_W$, $G$, $G_{SS}^{(1)}$, and MZI extinction ratio (extinction ratio due to non-ideal couplers), $ER$ (see Supplementary Material). \par
\begin{equation}
\begin{gathered}
    \frac{1}{G}=12\frac{\Delta T_{\pi}}{P_{\pi}}=\frac{6\lambda_0}{P_{\pi}L\gamma}\\
    C_W=\frac{5}{2}G\tau\\
    \frac{1}{G_{SS}^{(1)}}\approx\frac{P_{\pi}^{(1)}}{4G^2\Delta T_{\pi}}=\frac{P_{\pi}^{(1)}L\gamma}{2G^2\lambda_0}\\
    ER=\frac{P_{\mathrm{min}}}{P_{\mathrm{max}}}
\end{gathered}
\end{equation}
where $P_{\pi}^{(1)}$ is the power required to drive a TOPS adjacent to the main TOPS to induce $\pi$ phase shift at the main TOPS and $P_{\mathrm{min}}, P_{\mathrm{max}}$ correspond to the observed minimum and maximum optical powers, respectively. $C_W$, $G$, $G_{SS}^{(1)}$, and $ER$ can all be characterized experimentally.
\section{Row-Column TOPS Arrays}

\begin{figure}[t!]%
    \centering
    \subfloat[\centering]{{\includegraphics[height=0.361\textheight]{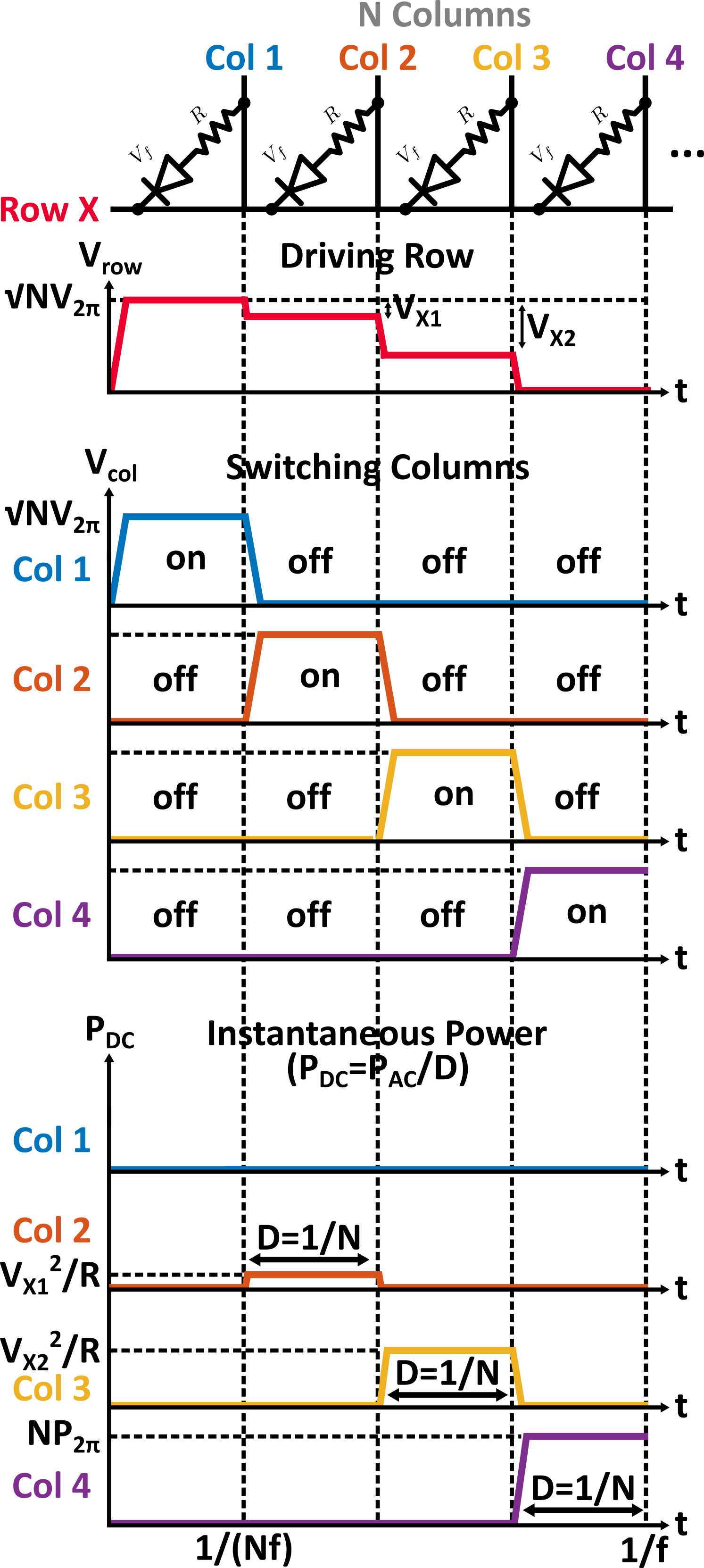} }}%
    \qquad\hspace{-25pt}
    \subfloat[\centering]{{\includegraphics[height=0.361\textheight]{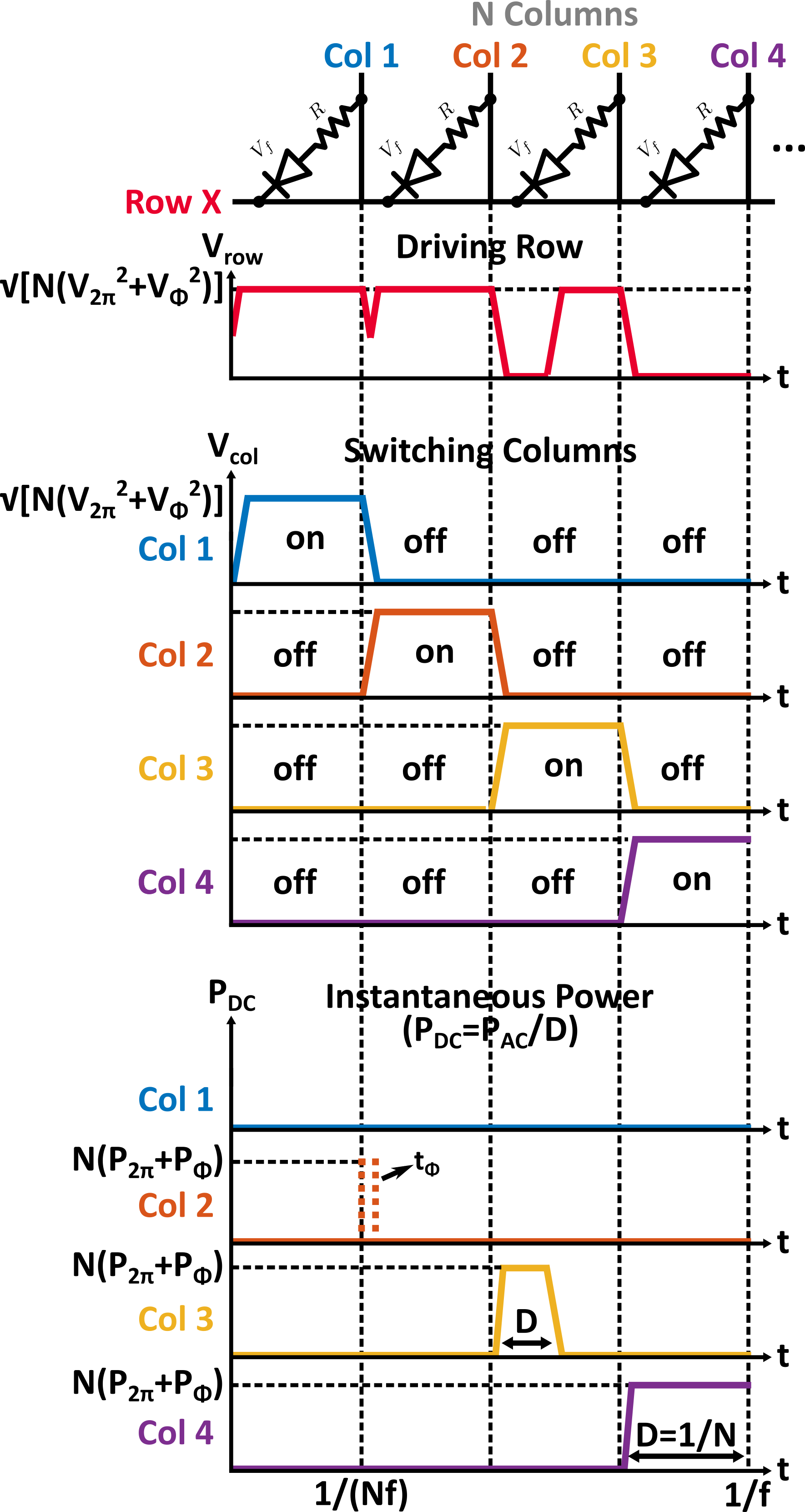} }}%
    \caption{Electrical waveforms for a) PAM and b) PWM driving of row-column TOPS arrays over one cycle. Columns are used to switch between different TOPS in the array while rows are used to drive the array. $\Delta V=V_{\mathrm{col}}-V_{\mathrm{row}}$ determines the instantaneous DC power ($P_{DC}$) dissipated in each TOPS. The average AC power dissipated ($P_{AC}$) is determined by the row voltage amplitude for PAM and by the row duty cycle for PWM. DC power can be calculated from row and column voltages via $P_{DC}=\Delta V^2/R$, where $R$ is the heater resistance. Also note that for PWM, there are additional terms (i.e. $P_\Phi, V_\Phi$) that correspond to the minimum pulse width ($t_\Phi$) resolvable by the limited bandwidth of the row driver.}
    \label{fig:PAMPWMwave}
\end{figure}

A row-column TOPS array can be realized by arranging multiple TOPS with diode switches in a row-column format. It can be driven with PAM \cite{Khachaturian2021, Khachaturian2021b} or PWM \cite{Ribeiro2017, Fatemi2019, Ribeiro2020} as seen in Fig. \ref{fig:PAMPWMwave}.\par

\begin{figure}[h!]
    \centering
    \includegraphics[width=\textwidth]{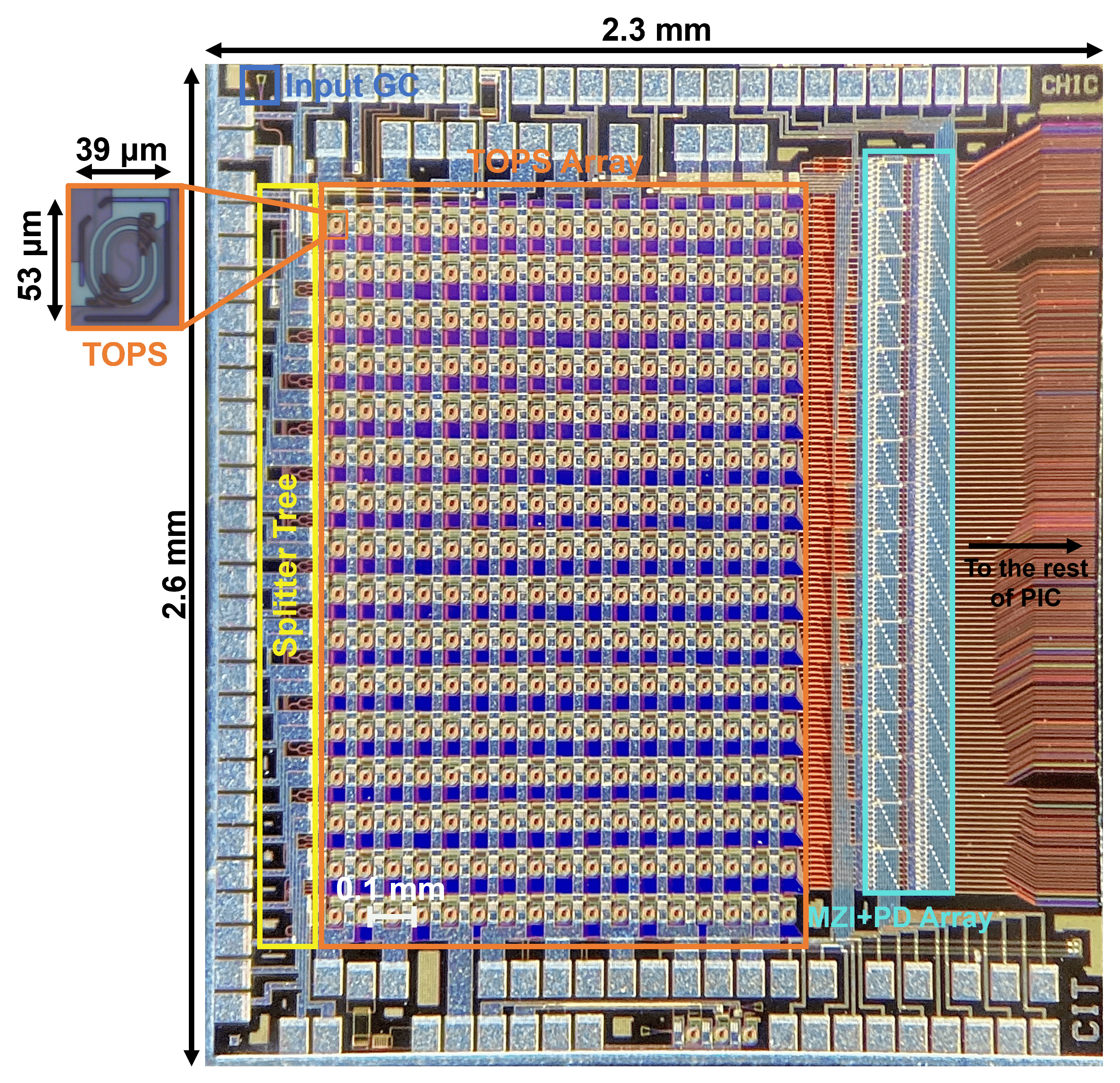}
    \caption{Micrograph of the realized 288-element TOPS array with feedback MZI+PD array.}
    \label{fig:chip}
\end{figure}

We realize a 32 $\times$ 9 row-column TOPS array on a silicon photonics platform as seen in Fig. \ref{fig:chip}. The row-column array utilizes a folded architecture, fitting two electrical rows per one physical row with a shorter optical path length, reducing the optical loss and increasing the coherence between branches. This architecture also reduces the electrical length of columns by half that allows a smaller metal width for routing, improving scalability.\par

\begin{figure}[t!]%
    \centering
    \subfloat[\centering]{{\includegraphics[width=\columnwidth]{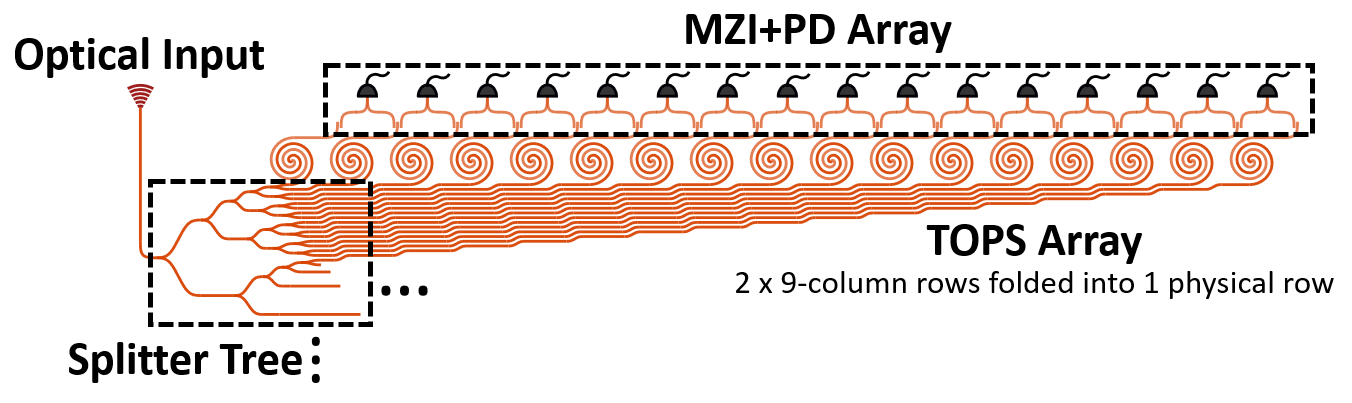} }}%

    \subfloat[\centering]{{\includegraphics[width=\columnwidth]{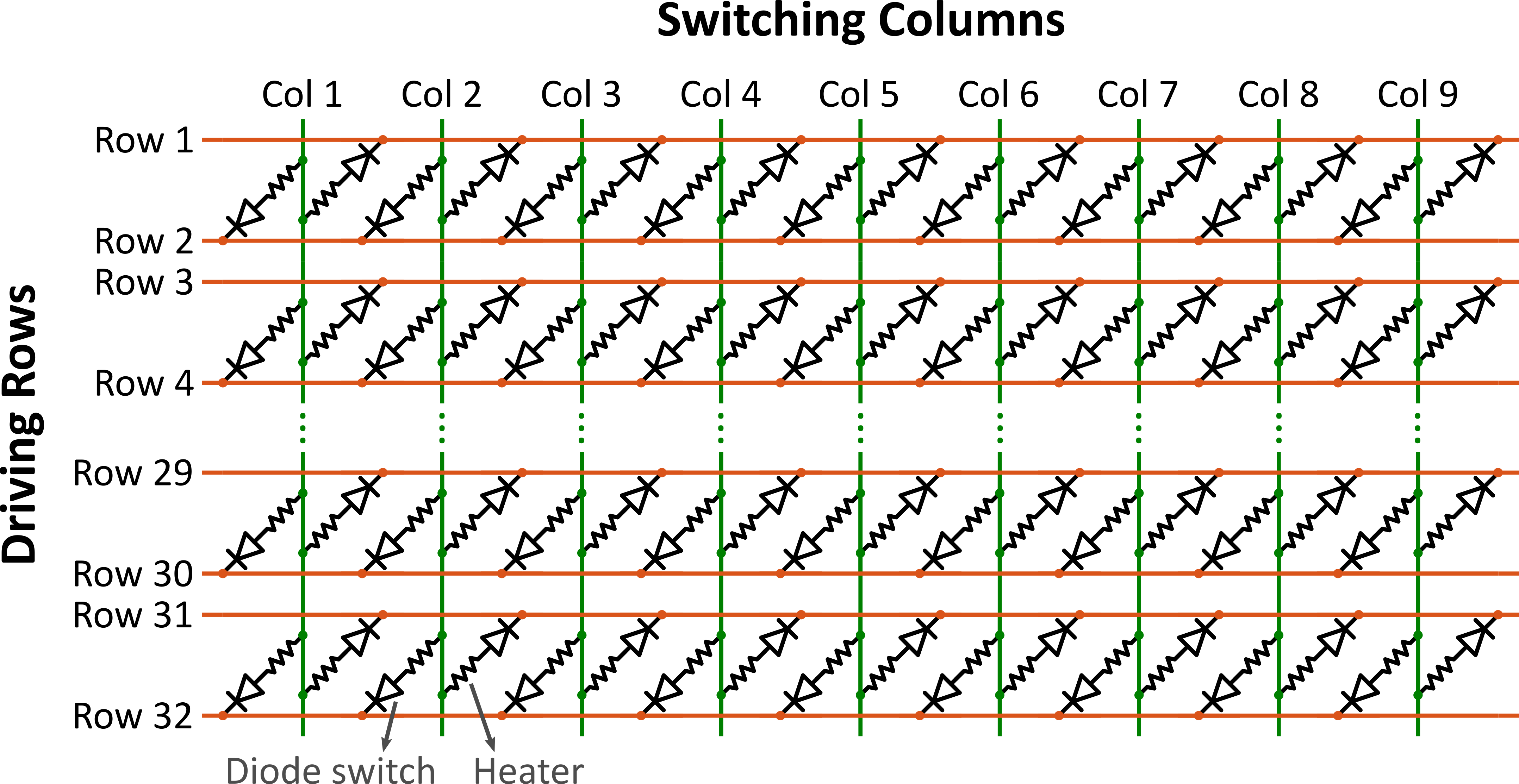} }}%
    \caption{a) Optical and b) electrical circuit diagrams of the TOPS array.}
    \label{fig:TOPSarray}
\end{figure}

Light is coupled to the chip via a grating coupler with 3.5 $\mathrm{dB}$ insertion loss and is distributed to the TOPS array via a 1:256 splitter tree with 2.4 $\mathrm{dB}$ excess loss. There are 32 heaters without any optical waveguides acting as dummy phase shifters \cite{Fatemi2019}. Each TOPS is comprised of a 300 $\mathrm{\mu}$m spiral waveguide surrounded by doped silicon serving as a Joule heater with an area of 1320 $\mathrm{\mu m^2}$. The heater is fitted with a silicon diode in series to switch on/off the TOPS in the array. There is also an oxide etch around each TOPS to minimize thermal crosstalk. TOPS is designed to have, on average, a $P_{\pi}$ of 10 $\mathrm{mW}$, a bandwidth of 30 $\mathrm{kHz}$, and an insertion loss of \textless0.2 $\mathrm{dB}$. To characterize the phase shifts from TOPS and use them as feedback information, we also realize an MZI with each TOPS pair that taps the TOPS outputs with 95:5 couplers and combines them with 0.3 $\mathrm{dB}$ excess loss. The output is fed to a PD in a row-column PD array. The number of electrical connections for the PD array is also reduced from $N^2$ to $N$ with the same row-column addressing scheme by adding a silicon diode switch in series with the PD \cite{Khachaturian2021c}. The total die area with the splitter tree, MZI+PD array, and electrical pads is 6 $\mathrm{mm^2}$. The total excess loss for each channel is 6.5 $\mathrm{dB}$. Circuit diagrams of the TOPS array are seen in Fig. \ref{fig:TOPSarray}.\par

\begin{figure}[t!]%
    \centering
    \subfloat[\centering]{{\includegraphics[width=0.333\columnwidth]{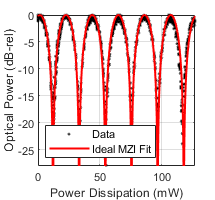} }}%
    \qquad\hspace{-31pt}
    \subfloat[\centering]{{\includegraphics[width=0.333\columnwidth]{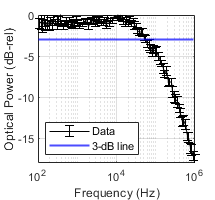} }}%
    \qquad\hspace{-31pt}
    \subfloat[\centering]{{\includegraphics[width=0.333\columnwidth]{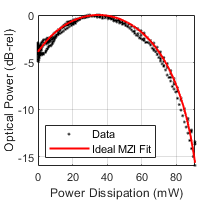} }}%
    \caption{Parameter extraction of a) $P_{\pi}$, b) $\tau$, and c) $P_{\pi}^{(1)}$.}%
    \label{fig:TOPScharac}%
\end{figure}

We characterize a TOPS test structure for the simulation parameters as shown in Fig. \ref{fig:TOPScharac}. We find its $P_{\pi}$ to be 10.6 $\mathrm{mW}$ and its bandwidth to be 56.8 $\mathrm{kHz}$. These parameters correspond to $C_W=440$ $\mathrm{pJ/K}$ and $G=G_{SW}=G_{SH}=62.8$ $\mathrm{\mu W/K}$, which are within an order of magnitude of the expected parameters from material properties given 0.5 $\mathrm{\mu m}$ waveguide width, 300 $\mathrm{\mu m}$ waveguide length, 220 $\mathrm{nm}$ waveguide height, and 2 $\mathrm{\mu m}$ BOX layer height ($C_W=54.7$ $\mathrm{pJ/K}$ with Si specific heat capacity of 20 $\mathrm{J/mol\cdot K}$ \cite{Chase1998}
and $G=71$ $\mathrm{\mu W/K}$ with $\mathrm{SiO_2}$ thermal conductivity of 1.1 $\mathrm{W/mK}$ \cite{Kleiner1996}%
). To characterize crosstalk, we drive a dummy phase shifter adjacent to the main TOPS without driving the main TOPS and find a $P_{\pi}^{(1)}$ of $62.6$ $\mathrm{mW}$ corresponding to $G_{SS}^{(1)}=4.57$ $\mathrm{\mu W/K}$. Furthermore, we characterize the extinction ratio to be $ER=-22.8$ $\mathrm{dB}$.\par
Due to fabrication variations, the characterized values from this parameter extraction can slightly differ from one TOPS to another in the array. Furthermore, TOPS placement in the array and thermal properties of the experimental setup may change these numbers. Therefore, we usually use the characterized parameters or simulated design specifications only as a starting point to fit the simulation parameters to the measured data. We also note that, as seen in Fig. \ref{fig:TOPScharac}b, the single TOPS frequency response is of first order, making the first-order approximation in our simulation valid within the frequency range we take measurements.\par
For an arbitrary array with $M$ rows and $N$ columns such as in Fig. \ref{fig:PAMPWMwave}, there are $M\times N$ phase shifters. For PAM, the duty cycle of each phase shifter is fixed at $1/N$. For PWM, the duty cycle of each phase shifter is varying but has a maximum of $1/N$. We also define a DC power ($P_{DC}$) that corresponds to the average AC power dissipated in the TOPS ($P_{AC}$). Then, for PAM, the DC power is varying but has a maximum of $N\times P_{2\pi}$, where $P_{2\pi}$ is the phase shifter power required for $2\pi$ phase shift. For PWM, the DC power is fixed at $N\times P_{2\pi}$. Therefore, $P_{DC}=P_{AC}/D$ for both PAM and PWM, where $D$ is the duty cycle.

\subsection{Control Electronics}
\begin{figure}[h!]
    \centering
    \includegraphics[width=\textwidth]{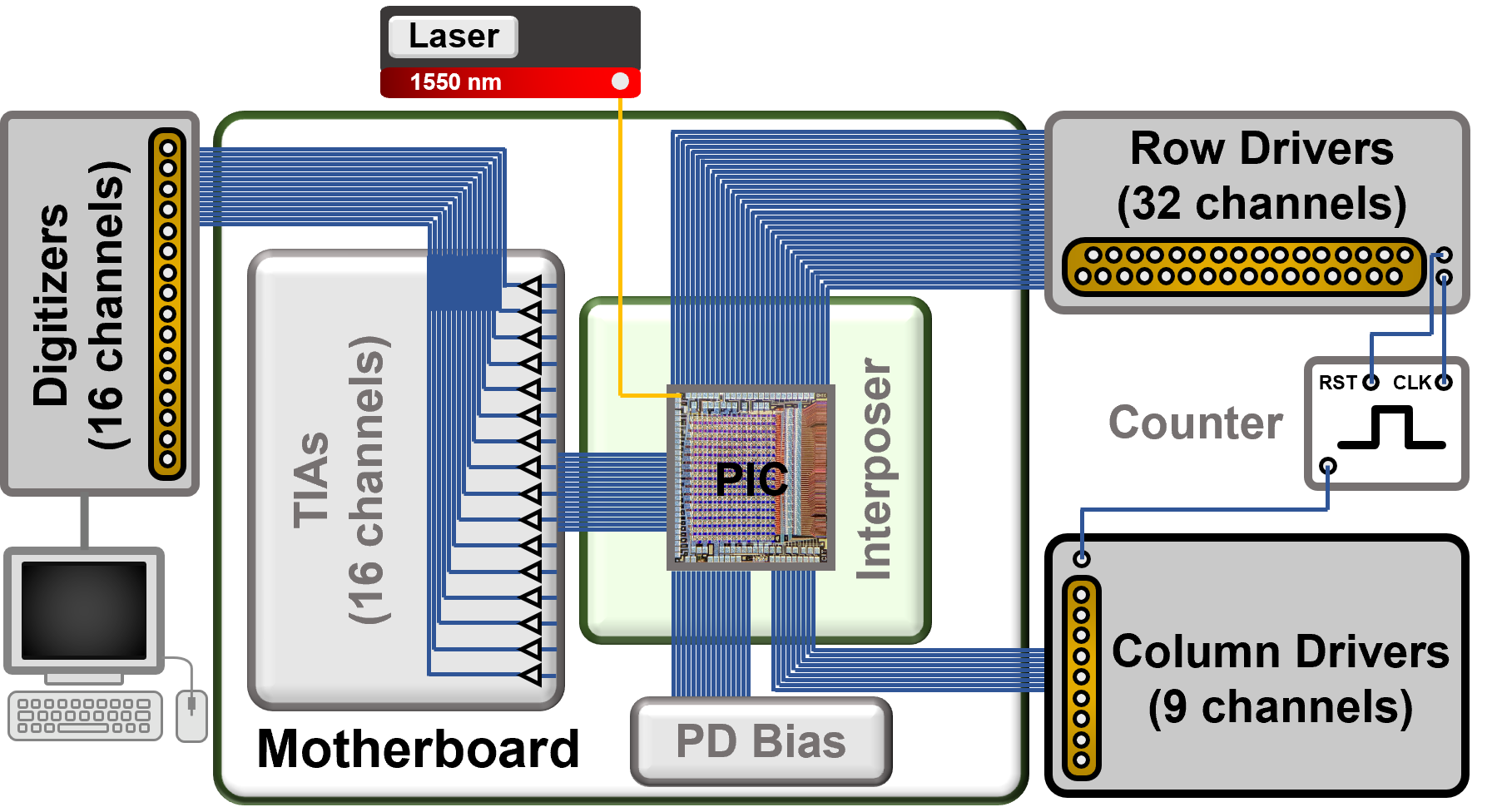}
    \caption{Driving and read-out electronics for the measurements.}
    \label{fig:setup}
\end{figure}

We first drive a single TOPS test structure with different waveforms to simulate row-column TOPS arrays of different sizes without thermal crosstalk and then the 32 $\times$ 9 TOPS array to investigate and correct thermal crosstalk. The control electronics for these measurements are shown in Fig. \ref{fig:setup}.\par
The PIC with the TOPS array is wirebonded to an interposer board sitting on top of a custom-made motherboard with bias circuitry and TIAs for 16 PD channels (16$\times$16 row-column multiplexed channels on-chip) and routing circuitry for driving 32 rows and 9 columns. PD channel multiplexing is achieved via a 4-to-16 demultiplexer, whose outputs are used to bias the PD rows, whereas PD columns are connected to TIAs for read-out. TIA outputs are digitized with 16 channels of 100 MSa/s, 14-bit digitizers. The switching columns are driven by a custom-made board with a 4-to-16 demultiplexer and operational amplifiers for 9 channels. The driving rows are driven by 32 channels of 1 GSa/s, 14-bit arbitrary waveform generators (AWGs) with an on-board FPGA. Row and column drivers share the same ground and are time-synchronized. Time synchronization is accomplished by a 4-bit synchronous counter with reset functionality that drives the logic inputs for the column driving demultiplexer. Desired waveforms for the 32 rows are pre-loaded to the row driver AWGs, and a separate waveform acting as the clock for the counter is sent to the counter board. Between each cycle, AWG also sends a trigger signal to the counter's reset pin to reset the switching cycle. To reduce noise coupled from the digital circuitry, driving and read-out electronics have two separate grounds with appropriate filtering for all the DC power supply.\par

We now investigate both PAM and PWM driving schemes in more detail.\par

\subsection{Pulse Amplitude Modulation}

\begin{figure}[t!]%
    \centering
    \subfloat[\centering]{{\includegraphics[width=0.5\columnwidth]{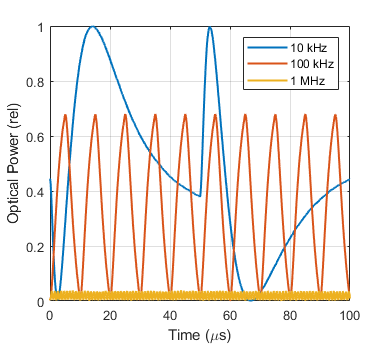} }}%
    \qquad\hspace{-31pt}
    \subfloat[\centering]{{\includegraphics[width=0.5\columnwidth]{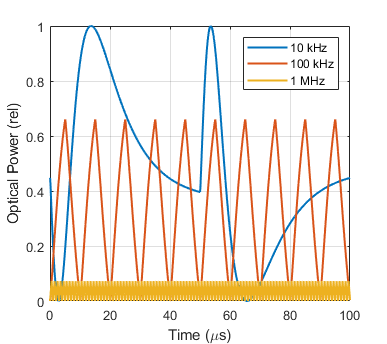} }}%
    \caption{a) Measured and b) simulated time-domain responses of PAM-driven TOPS at $P_\pi$. Since TOPS is driven at $2P_{\pi}$ DC power with 1/2 duty cycle, there is an observed phase swing around $\pi$. This phase swing is reduced at higher frequencies, leading to a higher extinction ratio.}
    \label{fig:PAMtime}
\end{figure}

PAM driving entails controlling the average power dissipated in the heaters by tuning the amplitude of the square waveform that has a constant pulse width as observed by a single TOPS. First, without crosstalk, we drive a single TOPS ($P_\pi=13.4$ $\mathrm{mW}$, $\tau=20.9$ $\mathrm{\mu s}$, $ER=-21.3$ $\mathrm{dB}$) with this waveform at different frequencies for $N=2$ at $P_\pi$ as seen in Fig \ref{fig:PAMtime}. The electronics bandwidth of the electronics is 200 $\mathrm{MHz}$, which determines the cut-off frequency of the low-pass filter in the simulation.\par

\begin{figure}[t!]%
    \centering
    \subfloat[\centering]{{\includegraphics[width=0.5\columnwidth]{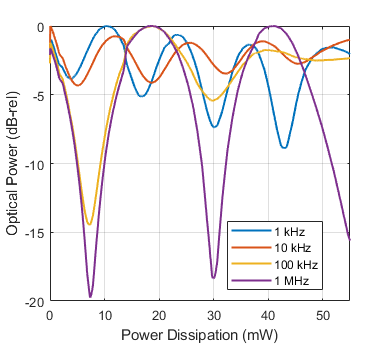} }}%
    \qquad\hspace{-31pt}
    \subfloat[\centering]{{\includegraphics[width=0.5\columnwidth]{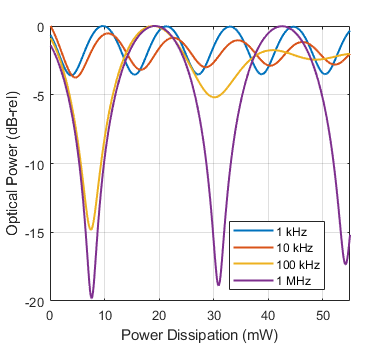} }}%
    \caption{a) Measured and b) simulated amplitude-domain responses (MZI curves) of PAM-driven TOPS. This shows the degradation of an ideal MZI response at lower driving frequencies for row-column TOPS arrays.}
    \label{fig:PAMamp}
\end{figure}

We then drive another single TOPS ($P_\pi=11.6$ $\mathrm{mW}$, $\tau=20$ $\mathrm{\mu s}$, $ER=-19.9$ $\mathrm{dB}$) at different frequencies with different AC powers to compare the simulated and experimental MZI curves seen in Fig. \ref{fig:PAMamp}. We note that at frequencies above 1 $\mathrm{MHz}$, extinction ratio measurement of TOPS is dominated by $ER$, extinction ratio same as that of a directly-addressed array.

The discrepancy between simulation and measurement at 1 $\mathrm{kHz}$ is due to the first approximation, through which we set $C_S$ and $G_{ES}$ significantly high, making the substrate a perfect heat sink. However, in measurement, substrate temperature rises a measurable amount at high AC powers.

\subsection{Pulse Width Modulation}
PWM driving entails controlling the average power dissipated in the heaters by tuning the duty cycle of the square waveform that has a constant amplitude as observed by a single TOPS. Without crosstalk, we again drive a single TOPS ($P_\pi=14.1$ $\mathrm{mW}$, $\tau=23.1$ $\mathrm{\mu s}$, $ER=-21.3$ $\mathrm{dB}$) with this waveform at different frequencies for $N=2$ at $P_\pi$ as seen in Fig. \ref{fig:PWMtime}.\par
We then again drive another single TOPS ($P_\pi=14.3$ $\mathrm{mW}$, $\tau=18.6$ $\mathrm{\mu s}$, $ER=-23.1$ $\mathrm{dB}$) at different frequencies with different AC powers to compare simulated and experimental MZI curves as seen in Fig. \ref{fig:PWMamp}.

\subsection{PAM vs. PWM}

With these measurements, we can compare PAM and PWM, whose waveforms and some of the definitions that will be used in the following analysis are shown in Fig. \ref{fig:PAMPWMwave}. PWM driving offers benefits in linearizing the control scheme and simplifying the electronics since voltage does not need to be tunable, whereas PAM requires tunability in amplitude. However, leveraging the thermal excitation model, a more detailed analysis shows more striking differences between PAM and PWM with respect to driving electronics specifications. Hence, we compare the two schemes regarding electronics' voltage swing and bandwidth.
\subsubsection{Voltage Swing}

\begin{figure}[t!]%
    \centering
    \subfloat[\centering]{{\includegraphics[width=0.5\columnwidth]{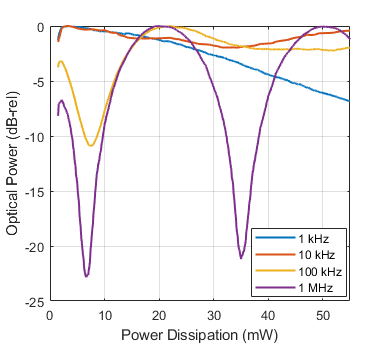} }}%
    \qquad\hspace{-31pt}
    \subfloat[\centering]{{\includegraphics[width=0.5\columnwidth]{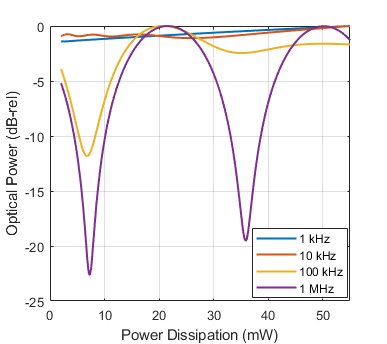} }}%
    \caption{a) Measured and b) simulated amplitude-domain responses (MZI curves) of PWM-driven TOPS. For PWM, the faster degradation of an ideal MZI response compared to PAM can be observed.}
    \label{fig:PWMamp}
\end{figure}

The voltage swing requirement for either scheme is set by the required phase shift range, which in most cases is $2\pi$ for full-wave coverage. For a single TOPS ($N$=1), the voltage swing requirement for PAM is $V_{2\pi}$, and $P_{2\pi}=V_{2\pi}^2/R$, where $R$ is the TOPS heater's electrical resistance. However, for PWM, since AC power is only tunable with duty cycle, an AC power that requires a pulse width corresponding to a bandwidth significantly larger than the electronics' bandwidth cannot be realized accurately. This can be observed in Fig. \ref{fig:PWMamp} at low AC powers (\textless 2 $\mathrm{mW}$). To realize phase shifts corresponding to these AC powers, voltage swing requirement needs to be $\sqrt{V_{2\pi}^2+V_{\Phi}^2}$, where $V_{\Phi}$ corresponds to the AC power required to recover the lost small phase shifts ($P_\Phi$). This couples the bandwidth and voltage swing requirements for PWM.\par
\begin{figure}[t!]%
    \centering
    \subfloat[\centering]{{\includegraphics[width=0.5\columnwidth]{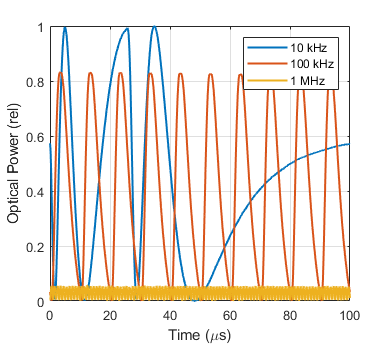} }}%
    \qquad\hspace{-31pt}
    \subfloat[\centering]{{\includegraphics[width=0.5\columnwidth]{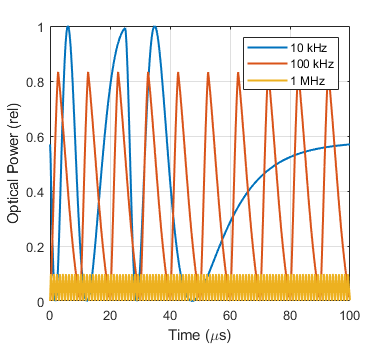} }}%
    \caption{a) Measured and b) simulated time-domain responses of PWM-driven TOPS at $P_\pi$. For PWM, since TOPS is driven at $4P_{\pi}$ DC power with 1/4 duty cycle, there is a phase swing around $\pi$ with a greater reduction in extinction ratio at lower driving frequencies compared to PAM.}
    \label{fig:PWMtime}
\end{figure}
Moreover, the voltage swing requirement scales linearly with $\sqrt{N}$, and for full-wave coverage, the driving electronics' output voltage swing should be $\sqrt{N}V_{2\pi}$ for PAM and $\sqrt{N(V_{2\pi}^2+V_{\Phi}^2)}$ for PWM. Therefore, the PAM voltage swing requirement is less stringent than that of PWM especially for large-scale TOPS arrays if $P_{\Phi}>P_{\mathrm{DAC}}$ where $P_{\Phi}$ is the power corresponding to $t_{\Phi}$ and $P_{\mathrm{DAC}}$ is the power corresponding to $t_{DAC}$, the sampling resolution of the row driver. This increased voltage swing requirement may become significantly detrimental to large-scale TOPS arrays with PWM driving since it also implies a larger reverse breakdown voltage requirement for the TOPS diodes and could extend the peak voltage requirement beyond the CMOS transistor breakdown voltage\cite{Fatemi2019,Miller2020}.\par
\subsubsection{Bandwidth}
\begin{figure}[h!]
    \centering
    \includegraphics[width=\textwidth]{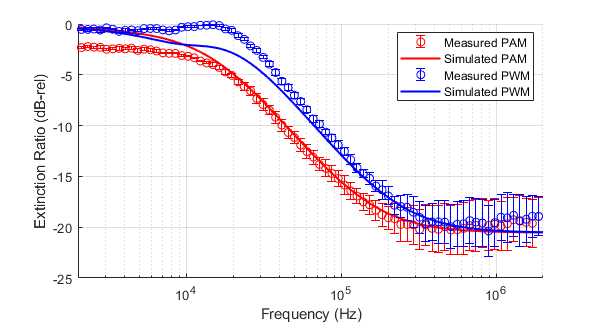}
    \caption{Measured and simulated frequency-domain responses of TOPS driven with PAM vs. PWM for $N=2$.}
    \label{fig:PAMPWMfreq}
\end{figure}

It can already be seen from the amplitude-domain responses that PAM driving reaches a higher extinction ratio at lower frequencies than PWM. To demonstrate this more clearly, we do a frequency-domain response comparison by finding the extinction ratio of the amplitude-domain response at each frequency for $N=2$. The resulting response obtained both experimentally and with the simulation is shown in Fig. \ref{fig:PAMPWMfreq}.\par
As seen here, for systems where bandwidth is the primary concern, PAM driving offers the benefit of reaching a higher extinction ratio than PWM driving. The reason for this can be explained by investigating Figs. \ref{fig:PAMtime} and \ref{fig:PWMtime} more carefully.\par
\begin{figure}[h!]
    \centering
    \includegraphics[width=\textwidth]{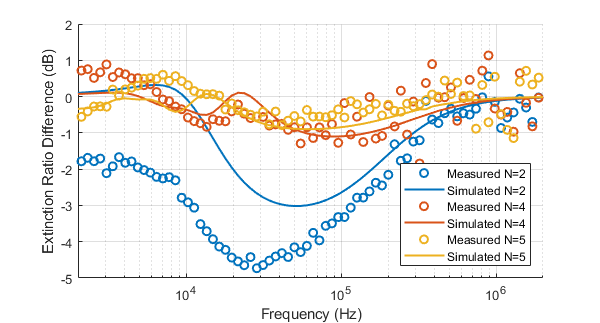}
    \caption{Measured and simulated differences in the extinction ratio for varying $N$ for TOPS driven with PAM vs. PWM.}
    \label{fig:PAMPWMdiff}
\end{figure}

In the time domain, TOPS phase shift oscillates back and forth centered around the intended phase shift. Phase swing of this oscillation determines the phase precision/extinction ratio of the TOPS/MZI. Phase swing is positively correlated with the time when TOPS is not driven, $t_{\mathrm{OFF}}$, the steady-state temperature difference of the waveguide when TOPS is driven vs. not driven, $\Delta T$, and is negatively correlated with the time constant, $\tau$, as determined by the solution to the system of ODEs in the model. While $\tau$ is specific to the TOPS design and $\Delta T$ is dictated by the required phase shift, $t_{\mathrm{OFF}}$ increases with pulse width, which is determined by electronics' bandwidth. PAM has an inherent advantage since its $t_{\mathrm{OFF}}$ is lower for a certain phase shift than that of PWM, $t_{\mathrm{OFF,PAM}}\leq t_{\mathrm{OFF,PWM}}$. The difference between these times, $\Delta t_{\mathrm{OFF}}=t_{\mathrm{OFF,PWM}}-t_{\mathrm{OFF,PAM}}$, determines the difference in the extinction ratio of these two schemes. Therefore, as the modulation frequency increases, $\Delta t_{\mathrm{OFF}}$ decreases, consequently reducing the difference in extinction ratio. The same effect takes place as $N$ is scaled, meaning both PAM and PWM bandwidth requirement scales linearly with $N$. To demonstrate this, we plot the difference in extinction ratio by performing the frequency-domain analysis in Fig. \ref{fig:PAMPWMdiff} for different $N$.\par
As seen in Fig. \ref{fig:PAMPWMdiff}, difference in the extinction ratio between PAM and PWM is minimal for large $N$ and for high bandwidth. However, this difference is significant (up to 4.7 $\mathrm{dB}$) for bandwidth-limited systems with small $N$ TOPS arrays.\par
Positive correlation between phase swing and $\Delta T$ also explains the reduction in the extinction ratio as power dissipation increases for higher phase shifts (e.g. $2\pi$) in Figs. \ref{fig:PAMamp} and \ref{fig:PWMamp}.\par
Finally, since voltage swing and bandwidth requirements are coupled for PWM, bandwidth could be used to minimize the voltage swing requirement by setting $P_{\Phi}<P_{\mathrm{DAC}}$, namely $t_{\Phi}<t_{DAC}$. However, since $t_{\Phi} \propto \frac{1}{BW}$, as we minimize the voltage swing requirement for PWM, bandwidth requirement becomes more stringent. Therefore, PWM suffers from a more stringent voltage swing requirement and/or a more stringent bandwidth requirement.\par
\section{Thermal Crosstalk}
\begin{figure}[h!]
    \centering
    \includegraphics[width=\textwidth]{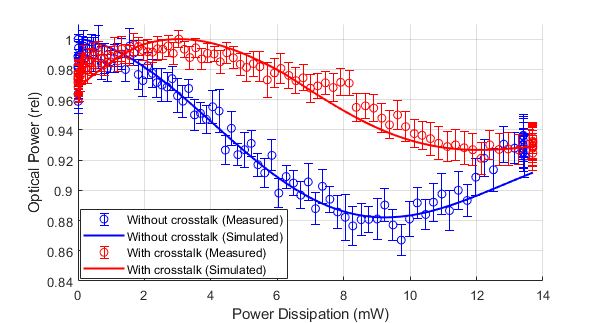}
    \caption{Measured and simulated MZI curves with and without crosstalk.}
    \label{fig:crosstalk}
\end{figure}

Now, we showcase the effects of thermal crosstalk in row-column TOPS arrays and demonstrate a correction algorithm with the information fed back from the integrated MZI+PD array and utilizing the previously detailed thermal excitation model. To maximize the effects of crosstalk, we select a TOPS in the middle of the array (row 16, column 5). To characterize the phase shift from the TOPS with and without crosstalk, we use the TOPS left of the main TOPS as the reference for the corresponding MZI. First, we drive the main TOPS with $N=9$ PAM of 200 $\mathrm{kHz}$ frequency (0.56 $\mu$s pulse width) but still keep the other TOPS off to have a baseline for the phase shift without crosstalk. Then, we drive 142 TOPS with the same waveform. We only keep the reference TOPS off since it serves as a reference and the TOPS right of the main TOPS off to introduce an asymmetry to observe the crosstalk effect. We drive the rest of the TOPS (16 rows x 9 columns) with a maximum amplitude of 9 V to maximize crosstalk. We also simulate the amplitude-domain response with the thermal excitation model. We use the simulation parameters ($C_W$=440 $\mathrm{pJ/K}$, $G$=62.8 $\mathrm{\mu W/K}$, $G_{SS}^{(1)}$=$4.57$ $\mathrm{\mu W/K}$, $ER$ = $-0.665$ $\mathrm{dB}$) extracted from the TOPS characterization as a starting point. We tune $G_{SS}^{(1)}=4.57$ $\mathrm{\mu W/K}$ to $G_{SS}^{(1)}=13.2$ $\mathrm{\mu W/K}$ since we neglected higher-order crosstalk in the simulation and crosstalk dynamics in the array are expected to be different from the test structure and from setup to setup. Hence, measured and simulated amplitude-domain responses for the main TOPS with and without crosstalk are shown in Fig. \ref{fig:crosstalk} with linear optical power on the y-axis. The measurement has a low extinction ratio due to stray light in the substrate reaching the integrated PD array but can be improved by adding Ge between components that absorbs stray light.\par
As seen here, thermal crosstalk causes a shift in the MZI curve. This shift consequently leads to a decrease in extinction ratio due to the aforementioned positive correlation between phase swing and $\Delta T$. \par
\subsection{Thermal Crosstalk Correction}
\begin{figure}[t!]%
    \centering
    \subfloat[\centering]{{\includegraphics[width=\columnwidth]{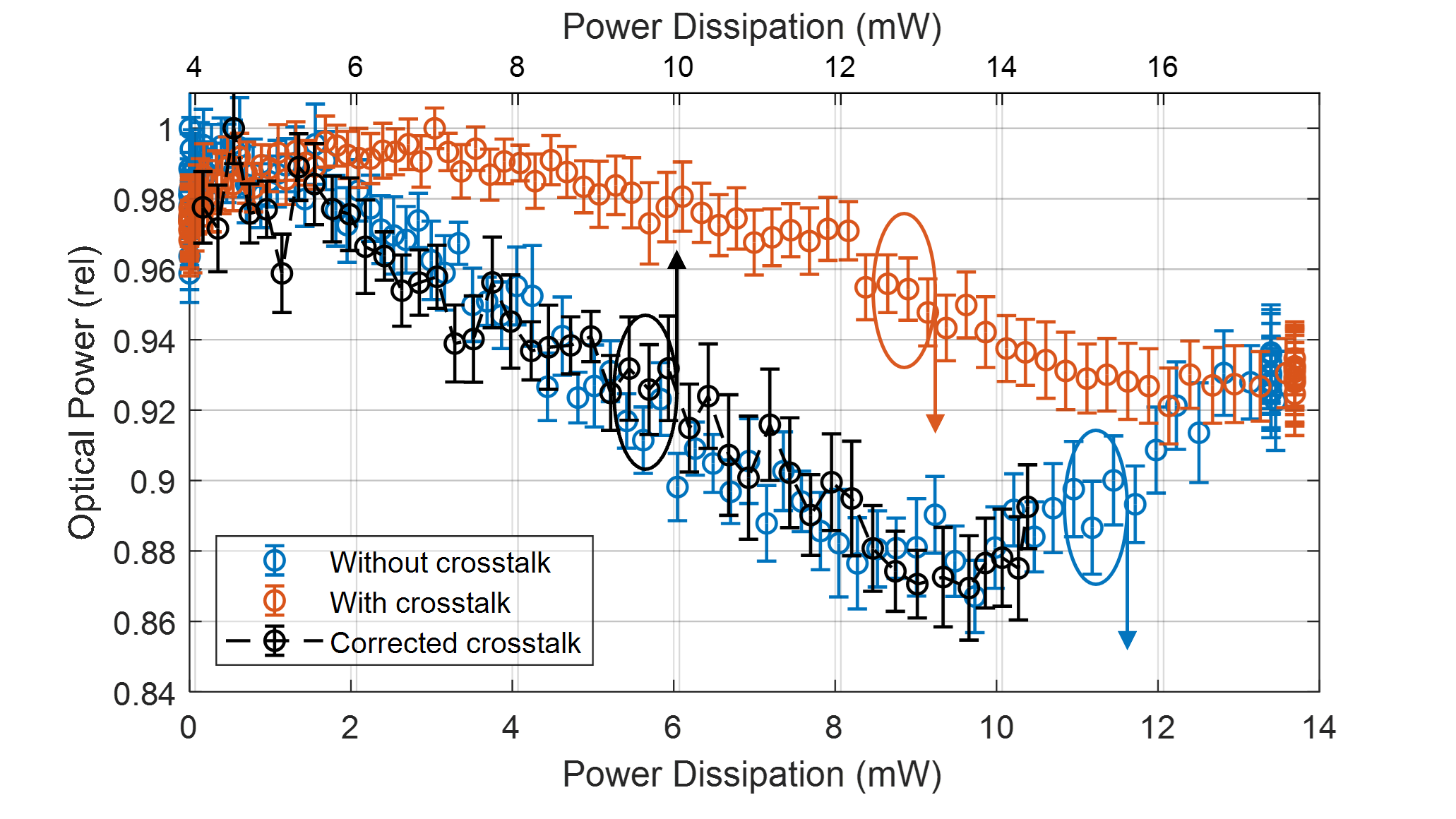} }}%
    \vspace{-1 pt}
    \subfloat[\centering]{{\includegraphics[width=\columnwidth]{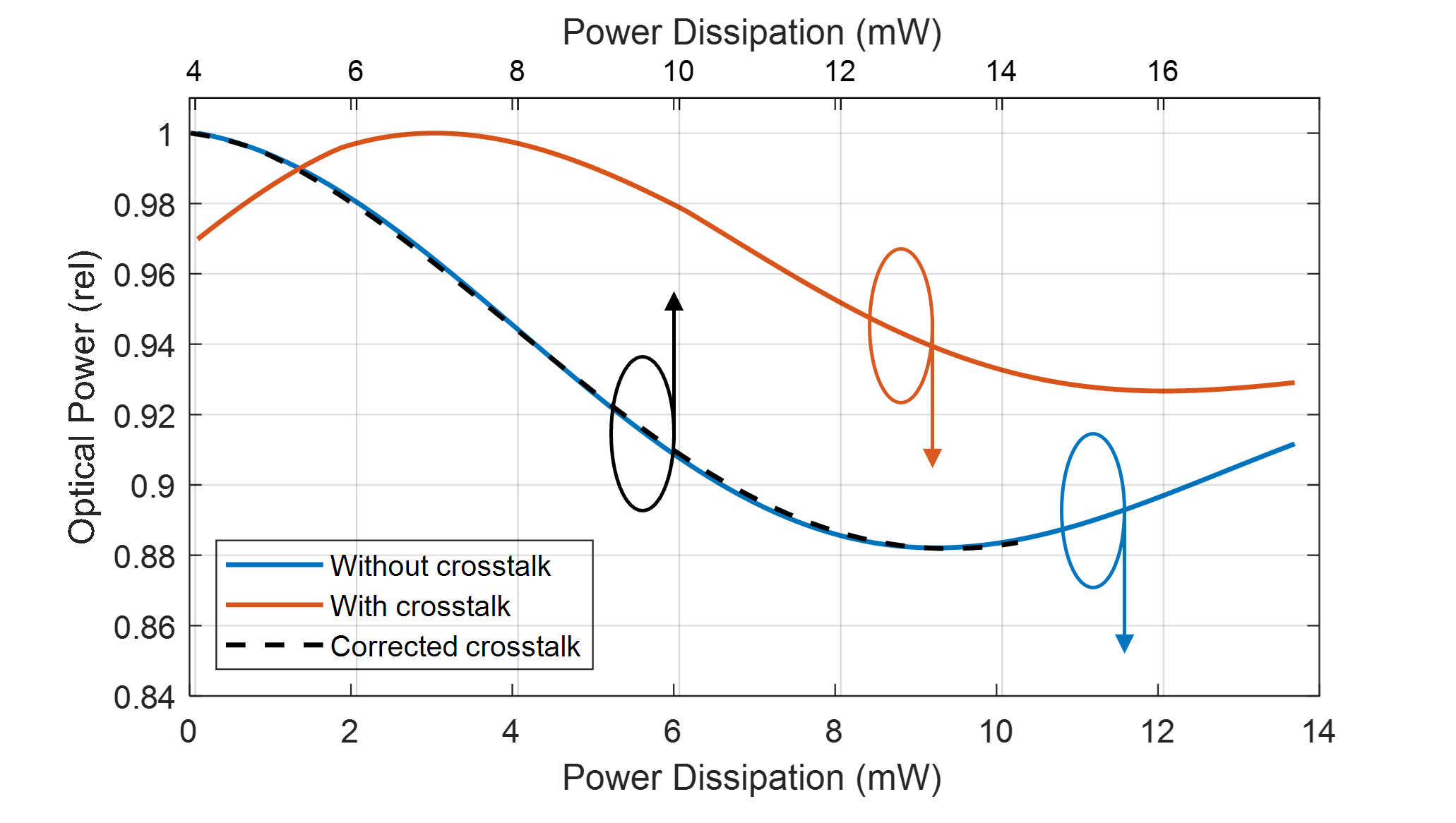} }}%
    \vspace{-7 pt}
    \caption{a) Measured and b) simulated MZI curves with, without crosstalk, and after crosstalk correction.}
    \label{fig:crosstalkcorrect}
\end{figure}

We now demonstrate a thermal crosstalk correction algorithm utilizing the thermal excitation model. For TOPS arrays with direct addressing, various thermal crosstalk correction algorithms have been proposed based on eigenmode decomposition \cite{Milanizadeh2019,Milanizadeh2020} and matrix inversion \cite{Harris2017}. These algorithms can also be implemented for row-column TOPS arrays to map the phase shifts from driving with crosstalk to driving without crosstalk. However, due to the phase swing caused by temporal multiplexing, the change in extinction ratio also needs to be corrected. In conjunction with our model, we implement an algorithm based on matrix inversion to correct thermal crosstalk. Assuming only first-order crosstalk and a linear system ($\frac{\partial \phi}{\partial P}=\mathrm{\mathbf{constant}}$), we define a square matrix, $C$, to specify the modulation and coupling coefficients at steady state. Then,
\begin{equation}
    \begin{bmatrix}
    \Phi_1\\
    \Phi_2\\
    \vdots
    \end{bmatrix}=\begin{bmatrix}
    C_{11} & C_{12} & \dots\\
    C_{12} & C_{22} & \dots\\
    \vdots & \vdots & \ddots
    \end{bmatrix}\begin{bmatrix}
    P_1\\
    P_2\\
    \vdots
    \end{bmatrix}    
\end{equation}
where $C_{ii+1}=\frac{\Gamma}{G^{(1)}}$ given $\Gamma = \frac{2\pi L\gamma}{\lambda_0}$ and $C_{ii+n}=0$ given $n>1$. Then, by finding the inverse of $C$, required $P$ can be calculated, $P=C^{-1}\Phi$. To correct the extinction ratio difference, we use the thermal excitation model to find the correct driving frequency. Simulation predicts a corrected driving frequency of 367 $\mathrm{kHz}$, setting the PAM driving frequency of the crosstalk corrected measurement. Measured and simulated MZI curves throughout the crosstalk correction algorithm are shown in Fig. \ref{fig:crosstalkcorrect}.\par
Since we used the parameters from the parameter extraction, this shows how our algorithm can be used in any arbitrary row-column TOPS array using only the characterization values from its test structure with minimal tuning. This also shows that in the case where parameter extraction is inaccurate, a row-column array of integrated feedback PDs and MZIs can be used to find the corrected parameters of all orders for accurate matrix inversion and extinction ratio correction.
\section{Conclusion}
We realize a 32 $\times$ 9 TOPS array based on a folded row-column array architecture with PAM and PWM control schemes and a thermal excitation model to analyze and compare both schemes. By leveraging our model, we shine light on the design requirements for the driving electronics and observe the time-domain, amplitude-domain, and frequency-domain behavior of row-column TOPS arrays. This enables waveform engineering for more versatile row-column TOPS arrays with less stringent electronic voltage swing and bandwidth requirements while maintaining the scalability advantage (M+N drivers for M $\times$ N TOPS).\par
We also investigate the effects of thermal crosstalk in row-column TOPS arrays and implement a crosstalk correction algorithm to completely cancel the thermal crosstalk for a TOPS in the middle of our array. This demonstrates negating the effects of thermal crosstalk in row-column TOPS architectures, unlocking higher-density TOPS integration.\par
From PIC dimensions, it is estimated that up to more than 40,000 crosstalk-corrected TOPS can be realized on a single 30 $\mathrm{mm}$ $\times$ 30 $\mathrm{mm}$ reticle by only modularizing this design with a potential for further scaling with larger $N$ row-column TOPS and feedback MZI+PD arrays. To the best of our knowledge, this realization signifies the largest row-column TOPS array with independent phase control ever demonstrated, paving the way for larger-scale and higher-density error-corrected PICs.

\bibliographystyle{IEEEtran}
\bibliography{References}

\ifCLASSOPTIONcaptionsoff
  \newpage
\fi

\begin{IEEEbiography}[{\includegraphics[width=1in,height=1.25in,clip,keepaspectratio]{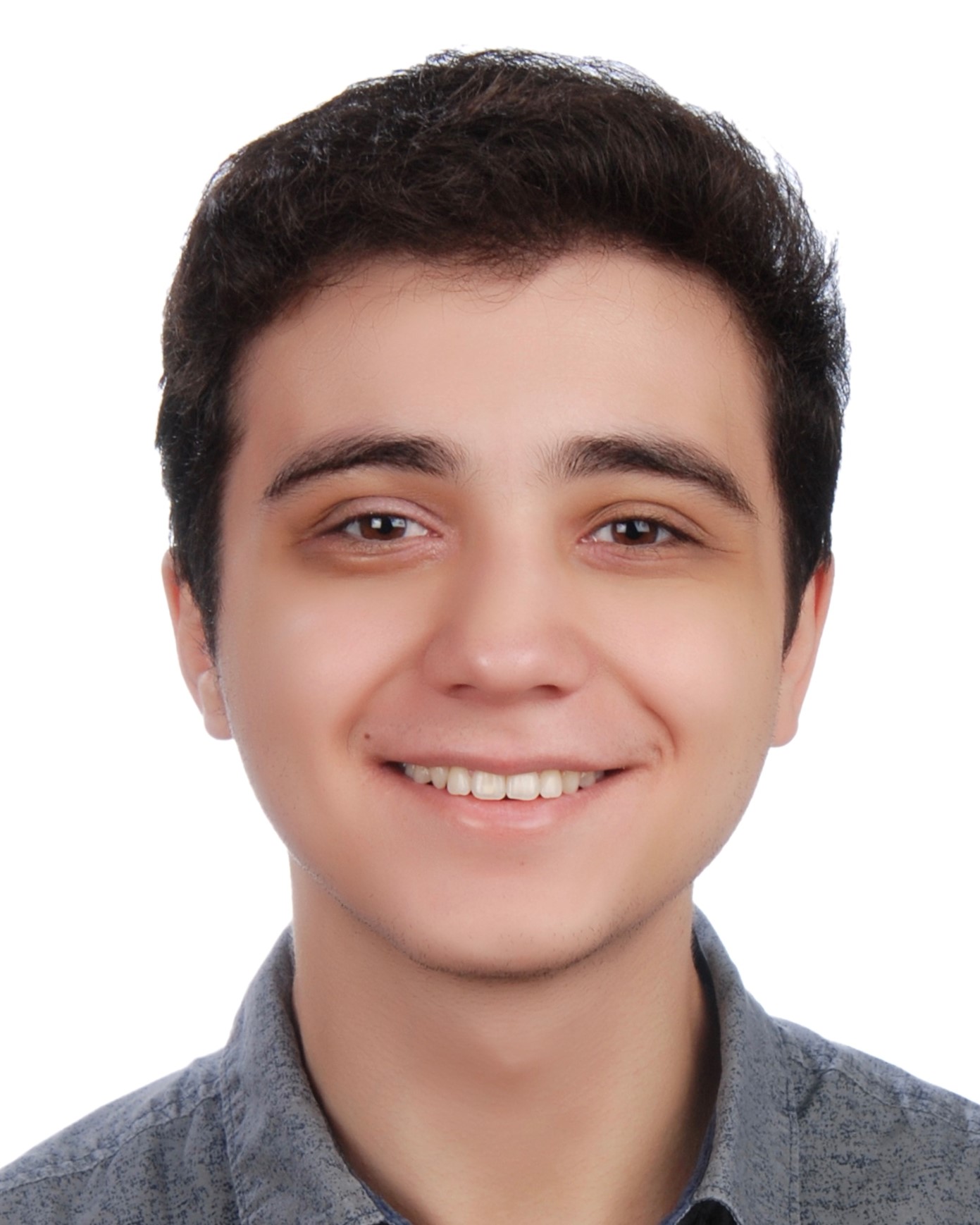}}]{B. Volkan Gurses} (Student Member, IEEE) received the B.S. degree in electrical engineering from the Georgia Institute of Technology, Atlanta, GA, USA, in 2020 and the M.S. degree in electrical engineering from the California Institute of Technology (Caltech), Pasadena, CA, USA, in 2022. He is currently pursuing the Ph.D. degree in electrical engineering at Caltech. He was awarded the R. David Middlebrook Fellowship, Caltech Engineering and Applied Sciences Division Fellowship, Tau Beta Pi Fellowship, and IEEE MTT-S Undergraduate/Pre-graduate Scholarship in 2020. His research interests are in large-scale photonic integrated circuits and silicon photonics for applications in sensing, machine learning, and quantum information processing.
\end{IEEEbiography}

\begin{IEEEbiography}[{\includegraphics[width=1in,height=1.25in,clip,keepaspectratio]{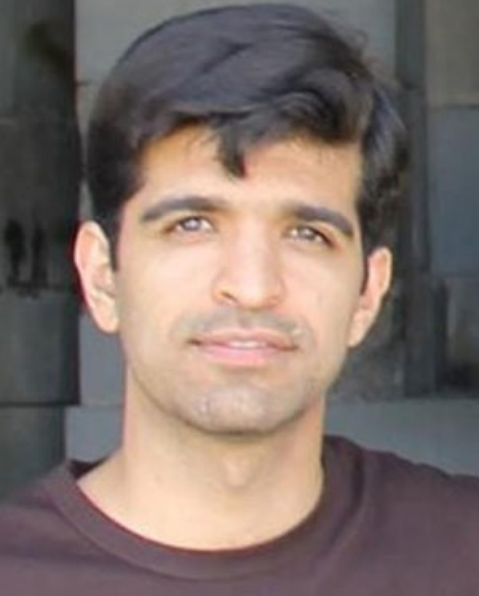}}]{Reza Fatemi} (Member, IEEE) received the B.S. degree in electrical engineering from K. N. Toosi University of Technology, Tehran, Iran, in 2011, the M.S. degree in electrical engineering from the Sharif University of Technology (SUT), Tehran, in 2013, and the Ph.D. degree in electrical engineering from the California Institute of Technology (Caltech), Pasadena, CA, USA, in 2020. He is currently a research scientist at Google Quantum. He was awarded the Oringer and Kilgore Fellowship in 2014 and the Analog Devices Outstanding Student Designer Award in 2015. His research interests are in silicon photonics and high-speed integrated circuit design.
\end{IEEEbiography}

\begin{IEEEbiography}[{\includegraphics[width=1in,height=1.25in,clip,keepaspectratio]{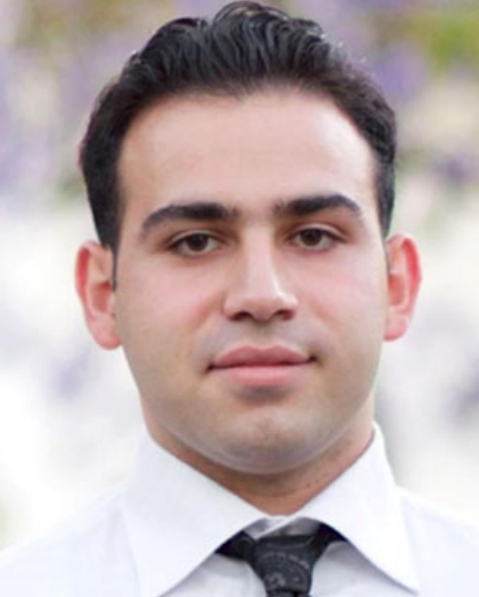}}]{Aroutin Khachaturian} (Member, IEEE) received the B.S., M.S., and Ph.D. degrees in electrical engineering from the California Institute of Technology in 2013, 2014, and 2020, respectively, where he is currently a Postdoctoral Scholar in electrical engineering. His research interests are in the design of integrated electro-optics systems for high-speed optical interconnects and photonic beamforming techniques for communications, imaging, and remote sensing. He was a recipient of the Killgore Fellowship and the Analog Devices, Inc., Outstanding Student Designer Award in 2013.
\end{IEEEbiography}

\begin{IEEEbiography}[{\includegraphics[width=1in,height=1.25in,clip,keepaspectratio]{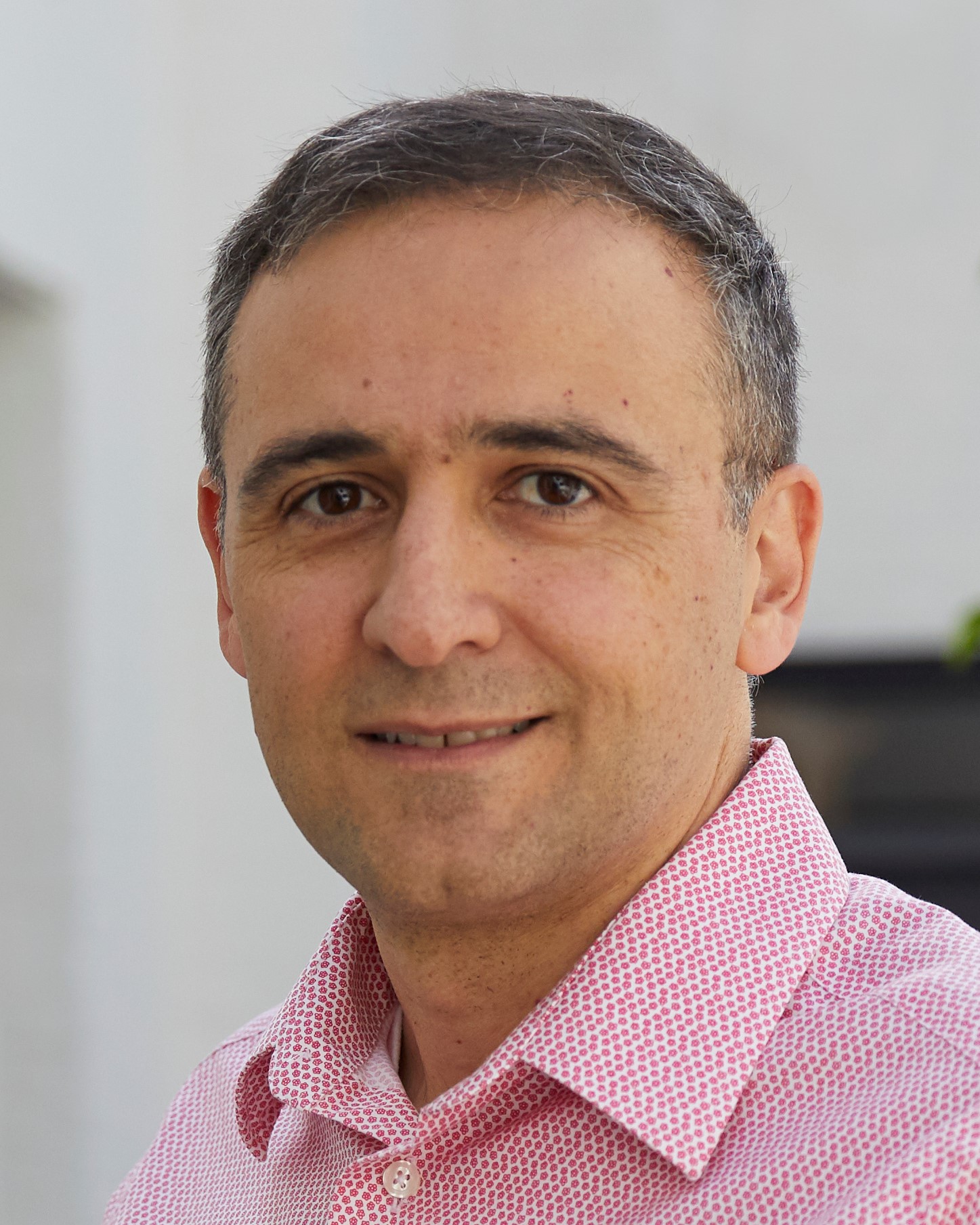}}]{Ali Hajimiri} (Fellow, IEEE) received the B.S. degree in electronics engineering from the Sharif University of Technology, and the M.S. and Ph.D. degrees in electrical engineering from Stanford University, Stanford, CA, USA, in 1996 and 1998, respectively.\par
He has been with Philips Semiconductors, where he worked on a BiCMOS chipset for GSM and cellular units from 1993 to 1994. In 1995, he was with Sun Microsystems working on the UltraSPARC microprocessors cache RAM design methodology. During the summer of 1997, he was with Lucent Technologies (Bell Labs), Murray Hill, NJ, USA, where he investigated low-phase-noise integrated oscillators. In 1998, he joined the Faculty of the California Institute of Technology, Pasadena, CA, USA, where he is the Bren Professor of Electrical Engineering and Medical Engineering and the Director of the Microelectronics Laboratory and the Co-Director of Space Solar Power Project. He has authored of Analog: Inexact Science, Vibrant Art (2020, Early Draft) a book on fundamental principles of analog circuit design and The Design of Low Noise Oscillators (Boston, MA, USA: Springer). He has authored and coauthored more than 250 refereed journal and conference technical articles and has been granted more than 130 U.S. patents with many more pending applications. His research interests are high-speed and high-frequency integrated circuits for applications in sensors, photonics, biomedical devices, and communication systems.\par
Prof. Hajimiri won the Feynman Prize for Excellence in Teaching, Caltech’s most prestigious teaching honor, as well as Caltech’s Graduate Students Council Teaching and Mentoring Award and the Associated Students of Caltech Undergraduate Excellence in Teaching Award. He was a co-recipient of the IEEE JOURNAL OF SOLID-STATE CIRCUITS Best Paper Award of 2004, the ISSCC Jack Kilby Outstanding Paper Award, the RFIC Best Paper Award, a two-time co-recipient of CICC Best Paper Award, and a three-time Winner of the IBM Faculty Partnership Award as well as the National Science Foundation CAREER Award and Okawa Foundation Award. He was the Gold Medal Winner of the National Physics Competition and the Bronze Medal Winner of the 21st International Physics Olympiad, Groningen, Netherlands. He was recognized as one of the top-ten contributors to International Solid-State Circuits Conference (ISSCC). In 2002, he co-founded Axiom Microdevices Inc., whose fully-integrated CMOS PA has shipped around 400,000,000 units, and was acquired by Skyworks Inc., in 2009. He has served on the Technical Program Committee of the ISSCC, as an Associate Editor of the IEEE JOURNAL OF SOLID-STATE CIRCUITS, IEEE TRANSACTIONS ON CIRCUITS AND SYSTEMS—PART II: EXPRESS BRIEFS, a member of the Technical Program Committees of the International Conference on Computer Aided Design, the Guest Editor of the IEEE TRANSACTIONS ON MICROWAVE THEORY AND TECHNIQUES, and the Guest Editorial Board of Transactions of Institute of Electronics, Information and Communication Engineers of Japan. He was selected to the TR35 top innovator’s list. He has served as a Distinguished Lecturer of the IEEE Solid-State and Microwave Societies. He is a Fellow of National Academy of Inventors.
\end{IEEEbiography}

\title{Supplementary Material for “Large-Scale Crosstalk-Corrected Thermo-Optic Phase Shifter Arrays in Silicon Photonics"}

\author{
B. Volkan Gurses,~\IEEEmembership{Student Member,~IEEE,} Reza Fatemi,~\IEEEmembership{Member,~IEEE,} Aroutin Khachaturian,~\IEEEmembership{Member,~IEEE,} and~Ali Hajimiri,~\IEEEmembership{Fellow,~IEEE}}

\maketitle

\IEEEpeerreviewmaketitle

\section{Matrix ODE Derivation for the Thermal Excitation Model}
\subsection{No Crosstalk}
\begin{figure}[h!]
    \centering
    \includegraphics[width=\textwidth]{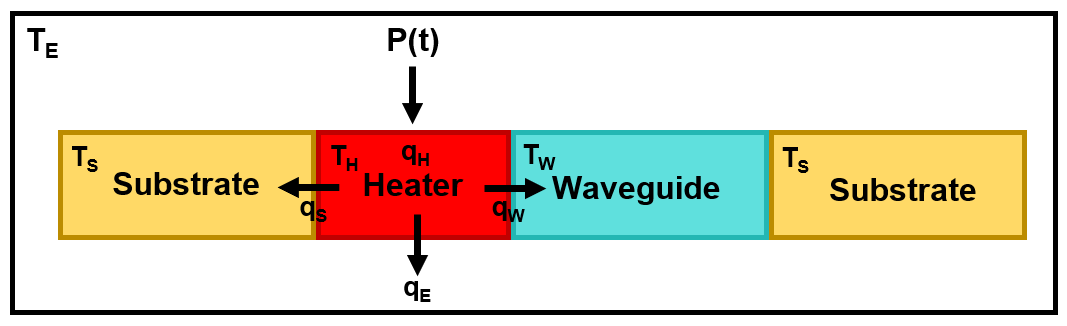}
    \caption{Lumped model for TOPS.}
    \label{fig:model}
\end{figure}
For no crosstalk, we have one TOPS with a heater, waveguide and substrate as shown in Fig. \ref{fig:model}. With the definitions detailed in Section IIA, we first write a PDE for the heater.
\begin{equation}
\begin{aligned}
    dP(t)&=\frac{1}{dt}\left(q_H+q_{W}+q_S+q_E\right)\\
    &=\rho_Hc_Hd V\frac{\partial}{\partial t} T(y,t)-k_{WH}d \vec{A}\cdot \nabla T(y,t)\\
    &-k_{SH}d \vec{A}\cdot \nabla T(y,t)-k_{EH}d \vec{A}\cdot \nabla T(y,t)
    \label{eq:heatpartial}
\end{aligned}
\end{equation}
Since temperature gradient is uniform over the surface area, $d \vec{A} \cdot \nabla T(y,t)=d A \nabla T(y,t)$. Then,
\begin{equation}
\begin{aligned}
     \frac{d P(t)}{d V}&=\rho_Hc_H\frac{\partial}{\partial t} T(y,t)-k_{WH} \nabla T(y,t)\frac{1}{d y}\\
     &-k_{SH} \nabla T(y,t)\frac{1}{d y}-k_{EH}\nabla T(y,t)\frac{1}{d y}
\end{aligned}
\end{equation}
Integrating over volume,
\begin{equation}
\begin{aligned}
    \int_V\left(\frac{d P(t)}{d V}\right)d V&=\int_V \rho_Hc_H\frac{\partial T(y,t)}{\partial t}d V\\
    &-\int_{A_{WH}} k_{WH}\nabla T(y,t)d A\\
    &-\int_{A_{SH}} k_{SH}\nabla T(y,t)d A\\
    &-\int_{A_{EH}} k_{EH}\nabla T(y,t)d A
\end{aligned}
\end{equation}
where we used $dV=dxdydz$, and $dA=dxdz$. Then,
\begin{equation}
\begin{aligned}
    P(t)&=C_H\frac{\partial T(y,t)}{\partial t}\\
    &-A_{WH}k_{WH}\nabla T(y,t)\\
    &-A_{SH}k_{SH}\nabla T(y,t)\\
    &-A_{EH}k_{EH}\nabla T(y,t)
\end{aligned}
\end{equation}
where $C_H$ is the heat capacity of the heater. Simplifying,
\begin{equation}
\begin{aligned}
    P(t)&=C_H\frac{\partial T(y,t)}{\partial t}-(G_{WH})\frac{\partial T(y,t)}{\partial y}\\
    &-(G_{SH})\frac{\partial T(y,t)}{\partial y}-(G_{EH})\frac{\partial T(y,t)}{\partial y}
    \label{eq:heaterdiff}
\end{aligned}
\end{equation}
where we took $G_{ij}=A_{ij}k_{ij}$ and noted that $A$ and $k$ are the same physical parameters as in (5). (\ref{eq:heaterdiff}) describes the process through which the applied electrical power is expended to increase the temperature of the heater, which then drives the heat exchange between the heater, the waveguide, and the substrate. The power exchanged with the waveguide is again expended through three mechanisms (again ignoring dissipated power in forms other than Joule heating):
\begin{enumerate}
    \item Heat used to increase the waveguide temperature
    \item Heat exchange with the substrate
    \item Heat exchange with surrounding environment other than the substrate
\end{enumerate}
Then,
\begin{equation}
\begin{aligned}
    C_W\frac{\partial T(y,t)}{\partial t}&=-G_{WH}\frac{\partial T(y,t)}{\partial y}-G_{SW}\frac{\partial T(y,t)}{\partial y}\\
    &-G_{EW}\frac{\partial T(y,t)}{\partial y}
    \label{eq:waveguidediff}
\end{aligned}
\end{equation}
where $C_W$ is the heat capacity of the waveguide and $G_{SW}$, $G_{EW}$ signifies the thermal coupling between the waveguide and the substrate, the environment respectively. Similarly, we can also derive a PDE for the substrate.
\begin{equation}
\begin{aligned}
    C_S\frac{\partial T(y,t)}{\partial t}&=-G_{SH}\frac{\partial T(y,t)}{\partial y}-G_{SW}\frac{\partial T(y,t)}{\partial y}\\
    &-G_{ES}\frac{\partial T(y,t)}{\partial y}
    \label{eq:subdiff}
\end{aligned}
\end{equation}
where $C_S$ is the heat capacity of the substrate and $G_{ES}$ signifies the thermal coupling between the substrate and the environment. Since we assumed uniform temperature distribution within all structures in all directions, spatial temperature distribution within each structure can be ignored, simplifying spatial partial derivatives to a lumped model. This is due to (for a linearly-varying spatial temperature distribution within each structure)
\begin{equation}
\begin{aligned}
    G_{ij}\frac{\partial T(y,t)}{\partial y}=A_{ij}k_i\frac{\Delta T_{i}}{\Delta y_{i}}+A_{ij}k_{ij}\Delta T_{ij}+A_{ij}k_j\frac{\Delta T_{j}}{\Delta y_{j}}
\end{aligned}
\end{equation}
where $\Delta T_i$ is the temperature difference within structure $i$, $\Delta T_{ij}$ is the temperature difference between structures $i$ and $j$ at their interface, $k_i$ is the thermal conductivity of the structure $i$, and $k_{ij}$ is the thermal contact conductivity between the structures $i$ and $j$. Considering only the heat flow at the interfaces between structures, $G_{ij}$ reduces to $A_{ij}k_{ij}$, where $k_{ij}$ is now the effective thermal contact conductivity between the respective structures. Defining $T_H$, $T_W$, $T_S$, and $T_E$ as lumped time-varying temperatures for the heater, waveguide, substrate and environment, respectively, we simplify (\ref{eq:heaterdiff}) to
\begin{equation}
\begin{aligned}
    P&=C_H\frac{d T_H}{d t}-(G_{WH})(T_W-T_H)-G_{SH}(T_S-T_H)\\
    &-G_{EH}(T_E-T_H)
    \label{eq:heatertemp}
\end{aligned}
\end{equation}
Similarly, (\ref{eq:waveguidediff}) becomes
\begin{equation}
\begin{aligned}
    C_W\frac{d T_{W}}{d t}&=G_{WH}(T_H-T_{W})+G_{SW}(T_S-T_{W})\\
    &+G_{EW}(T_E-T_{W})
    \label{eq:waveguidetemp}
\end{aligned}
\end{equation}
and (\ref{eq:subdiff}) becomes
\begin{equation}
\begin{aligned}
    C_S\frac{d T_S}{d t}&=G_{SH}(T_H-T_S)+G_{SW}(T_W-T_{S})\\
    &+G_{ES}(T_E-T_{S})
    \label{eq:subtemp}
\end{aligned}
\end{equation}
Hence, using using (\ref{eq:heatertemp}), (\ref{eq:waveguidetemp}), and (\ref{eq:subtemp}), we write the matrix ODE in (6) for this lumped system.
\subsection{With Crosstalk}
We now redo the above derivation including the crosstalk between all phase shifters. If we have N phase shifters, for phase shifter $i$, (\ref{eq:heaterdiff}) becomes
\begin{equation}
\begin{aligned}
    P_{i}(t)&=C_{H,i}\frac{\partial T(y,t)}{\partial t}-\sum_{j=1}^N \left(G_{H,iH,j}\frac{\partial T(y,t)}{\partial y}\right)\\
    &-\sum_{j=1}^N \left(G_{W,jH,i}\frac{\partial T(y,t)}{\partial y}\right)\\
    &-\sum_{j=1}^N \left(G_{S,jH,i}\frac{\partial T(y,t)}{\partial y}\right)-(G_{EH,i})\frac{\partial T(y,t)}{\partial y}
    \label{eq:heaterfull}
\end{aligned}
\end{equation}
where $S_i$, $W_i$ and $H_i$ are the substrate, waveguide, and heater of phase shifter $i$ respectively. (\ref{eq:waveguidediff}) becomes
\begin{equation}
\begin{aligned}
    C_{W,i}\frac{\partial T(y,t)}{\partial t}&=-\sum_{j=1}^N \left(G_{W,iH,j}\frac{\partial T(y,t)}{\partial y}\right)\\
    &-\sum_{j=1}^N \left(G_{W,iW,j}\frac{\partial T(y,t)}{\partial y}\right)\\
    &-\sum_{j=1}^N \left(G_{S,jW,i}\frac{\partial T(y,t)}{\partial y}\right)\\
    &-G_{EW,i}\frac{\partial T(y,t)}{\partial y} \label{eq:waveguidefull}
\end{aligned}
\end{equation}
and (\ref{eq:subdiff}) becomes
\begin{equation}
\begin{aligned}
    C_{S,i}\frac{\partial T(y,t)}{\partial t}&=-\sum_{j=1}^N \left(G_{S,iH,j}\frac{\partial T(y,t)}{\partial y}\right)\\
    &-\sum_{j=1}^N \left(G_{S,iW,j}\frac{\partial T(y,t)}{\partial y}\right)\\
    &-\sum_{j=1}^N \left(G_{S,iS,j}\frac{\partial T(y,t)}{\partial y}\right)\\
    &-G_{ES,i}\frac{\partial T(y,t)}{\partial y}
    \label{eq:subfull}
\end{aligned}
\end{equation}
Now again, considering uniform temperature distributions within all structures, we define $T_{H,i}$, $T_{W,i}$, $T_{S,i}$, and $T_E$ as the lumped time-varying temperatures for the heater, waveguide, substrate of TOPS $i$ and environment, respectively. (\ref{eq:heaterfull}) simplifies to
\begin{equation}
\begin{aligned}
P_{H,i}&=C_{H,i}\frac{d T_{H,i}}{d t}\\
    &-\sum_{j=1}^N
    \left[G_{H,iH,j}\left(T_{H,j}-T_{H,i}\right)\right]\\
    &-\sum_{j=1}^N
    \left[G_{W,jH,i}\left(T_{W,j}-T_{H,i}\right)\right]\\
    &-\sum_{j=1}^N\left[G_{S,jH,i}\left(T_{S,j}-T_{H,i}\right)\right]-G_{EH,i}\left(T_E-T_{H,i}\right)
    \label{heaterode}
\end{aligned}
\end{equation}
Similarly, (\ref{eq:waveguidefull}) becomes
\begin{equation}
\begin{aligned}
C_{W,i}\frac{d T_{W,i}}{d t}&=\sum_{j=1}^N\left[G_{W,iH,j}\left(T_{H,j}-T_{W,i}\right)\right]\\
&+\sum_{j=1}^N \left[G_{W,iW,j}\left(T_{W,j}-T_{W,i}\right)\right]\\
&+\sum_{j=1}^N\left[G_{S,jS,i}\left(T_{S,j}-T_{W,i}\right)\right]+G_{ES,i}\left(T_E-T_{W,i}\right)
\label{waveguideode}
\end{aligned}
\end{equation}
and (\ref{eq:subfull}) becomes
\begin{equation}
\begin{aligned}
    C_{S,i}\frac{d T_{S,i}}{d t}&=\sum_{j=1}^N
    \left[G_{S,iH,j}\left(T_{H,j}-T_{S,i}\right)\right]\\
    &+\sum_{j=1}^N
    \left[G_{S,iW,j}\left(T_{W,j}-T_{S,i}\right)\right]\\
    &+\sum_{j=1}^N\left[G_{S,jS,i}\left(T_{S,j}-T_{S,i}\right)\right]+G_{ES,i}\left(T_E-T_{S,i}\right)
    \label{subode}
\end{aligned}
\end{equation}
Using, (\ref{heaterode}), (\ref{waveguideode}), and (\ref{subode}), we write the generalized matrix ODE defined in (9).

\section{Derivation of the Parameter Extraction Equations}

\subsection{$G$ and $C_W$}

\begin{figure}[H]
    \centering
    \includegraphics[width=0.85\textwidth]{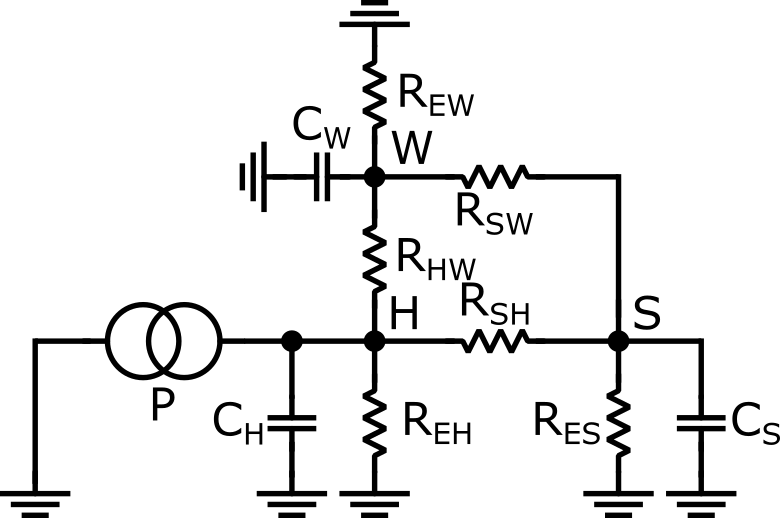}
    \caption{RC circuit model of a single TOPS.}
    \label{fig:RCgeneral}
\end{figure}
\begin{figure}[h!]
    \centering
    \includegraphics[width=0.4\textwidth]{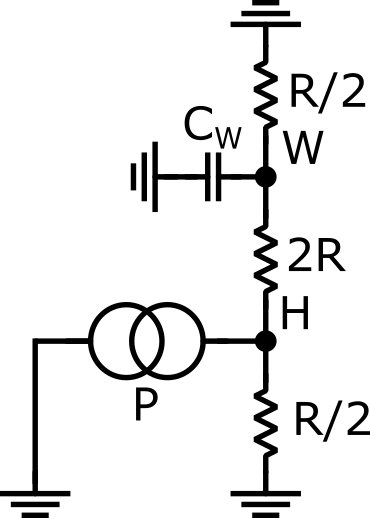}
    \caption{Reduced circuit model of a single TOPS with the introduced approximations.}
    \label{fig:RCsimple}
\end{figure}

To find the equations for $G$ and $C_W$, we draw the RC circuit model corresponding to the matrix ODE in (6) as shown in Fig. $\ref{fig:RCgeneral}$. With the approximations introduced in Section IIB ($G_{ES}\gg G_{SW}=G_{SH}=G$ and $C_S\gg C_W\gg C_H$), the reduced circuit model becomes as shown in Fig. \ref{fig:RCsimple}.\par
Neglecting $C_W$ for DC analysis, $\Delta T_\pi$ at the waveguide node can be calculated as
\begin{equation}
    \begin{aligned}
        \Delta T_\pi=\frac{P_\pi}{6}\frac{R}{2}
    \end{aligned}
    \label{deltaT}
\end{equation}
Rearranging (\ref{deltaT}),
\begin{equation}
    \begin{aligned}
    R=\frac{1}{G}=12\frac{\Delta T_{\pi}}{P_{\pi}}=\frac{6\lambda_0}{P_{\pi}L\gamma}
    \end{aligned}
\end{equation}
Now, with $C_W$,
\begin{equation}
    \begin{aligned}
        \Delta T_\pi=P_\pi\frac{R_{eq}}{R_{eq}+2R}
    \end{aligned}
\end{equation}
where $R_{eq}=\frac{R}{2}\parallel\frac{1}{j\omega C_W}=\frac{R}{2+j\omega RC_W}$. Then,
\begin{equation}
    \begin{aligned}
        \Delta T_\pi&=P_\pi\frac{R}{5R+2j\omega R^2C_W}\\
        &=P_\pi\frac{1}{5+j2\omega RC_W}
    \end{aligned}
\end{equation}
Since $\tau=1/\omega_{c}$,
\begin{equation}
    \begin{aligned}
        C_W&=\frac{5}{2\omega_c R}\\
        &=\frac{5\tau}{2R}=\frac{5}{2}G\tau
    \end{aligned}
\end{equation}
\subsection{$G_{SS}^{(1)}$}
\begin{figure}[h!]
    \centering
    \includegraphics[width=\textwidth]{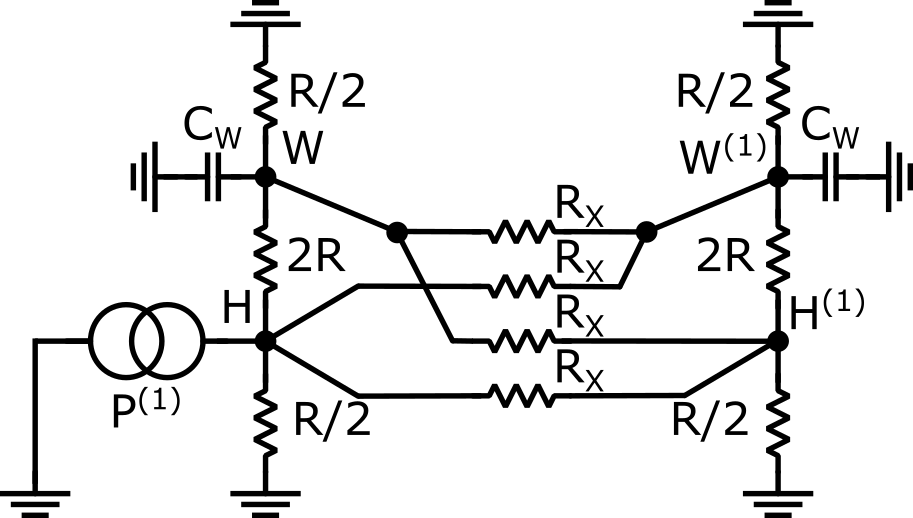}
    \caption{Reduced circuit model of a coupled TOPS pair of first-order crosstalk.}
    \label{fig:RCcoupled}
\end{figure}
To extract $G_{SS}^{(1)}$, we draw the circuit diagram for a coupled TOPS pair of first-order crosstalk, again with the same approximations, as shown in Fig. \ref{fig:RCcoupled}. Solving for the temperature at $W^{(1)}$ gives the following transcendental equation.
\begin{equation}
    \begin{aligned}
        \Delta T_{\pi}=P_\pi^{(1)}\frac{R^2\left(4 R^4 R_X+12 R^3 R_X^2+9 R^2 R_X^3\right)}{4 \left(8 R^5 R_X+28 R^4 R_X^2+30 R^3 R_X^3+9 R^2 R_X^4\right)}
    \end{aligned}
\end{equation}
where $R_X=2 R+R_{SS}^{(1)}$. This can be solved numerically for $R_{SS}^{(1)}$, or for an analytical expression, we can set $R_{SS}^{(1)}\gg R$ since $R_{SS}^{(1)}$ is expected to be much larger than $R$ for a good TOPS array design. With this approximation, last term in both the numerator and denominator dominates.
\begin{equation}
    \begin{aligned}
        \Delta T_{\pi}&\approx P_\pi^{(1)}\frac{9 R^4 R_X^3}{36 R^2 R_X^4}\\
        &=P_\pi^{(1)}\frac{R^2}{4R_X}\\
        &\approx P_\pi^{(1)}\frac{R^2}{4R_{SS}^{(1)}}
    \end{aligned}
\end{equation}
Hence,\begin{equation}
    \begin{aligned}
        R_{SS}^{(1)}=\frac{1}{G_{SS}^{(1)}}&\approx\frac{P_{\pi}^{(1)}R^2}{4\Delta T_{\pi}}\\
        &=\frac{P_{\pi}^{(1)}}{4G^2\Delta T_{\pi}}
    \end{aligned}
\end{equation}
\end{document}